\documentclass[twocolumn,amsmath,amssymb,prl]{revtex4}
\usepackage{latexsym}
\usepackage{graphicx}% Include figure files
\usepackage{color}
\usepackage{amsmath}

\begin{document}
\title{Missing Shapiro steps in topologically trivial Josephson Junction on InAs quantum well}

\author{Matthieu C. Dartiailh$^{1}$}
\author{Joseph J. Cuozzo$^{2}$}
\author{William~Mayer$^{1}$}
\author{Joseph~Yuan$^{1}$}
\author{Kaushini~S.~Wickramasinghe$^{1}$}
\author{Enrico~Rossi$^{2}$}
\author{Javad~Shabani$^{1}$}
\affiliation{$^{1}$Center for Quantum Phenomena, Department of Physics, New York University, NY 10003, USA\\
$^{2}$ Department of Physics, William \& Mary, Williamsburg, VA 23187, USA
}

\date{\today}

\begin{abstract}
    Josephson junctions hosting Majorana fermions have been predicted to exhibit a 4$\pi$ periodic current phase relation. The experimental consequence of this periodicity is the disappearance of odd steps in Shapiro steps experiments. Experimentally, missing odd Shapiro steps have been observed in a number of materials systems with strong spin-orbit coupling and have been interpreted in the context of topological superconductivity. Here, we report on missing odd steps in topologically trivial Josephson junctions fabricated on InAs quantum wells. We ascribe our observations to the high transparency of our junctions allowing Landau-Zener transitions. The probability of these processes is found to be independent of the drive frequency. We analyze our results using a bi-modal transparency distribution which demonstrates that only few modes carrying 4$\pi$ periodic current are sufficient to describe the disappearance of odd steps. Our findings highlight the elaborate circumstances that have to be considered in the investigation of the 4$\pi$ Josephson junctions in relationship to topological superconductivity.
\end{abstract}

\maketitle

Recently the drive to understand and control the order parameter characterizing the collective state of electrons in quantum heterostructures has intensified. New physical behavior can emerge that is absent in the isolated constituent materials \cite{Cao_Gr_2018}.  With regards to superconductivity this has opened a whole new area of investigation in the form of topological superconductivity \cite{fu_superconducting_2008, oreg_helical_2010,Lutchyn_MF_2010}.  Topological superconductors are expected to host Majorana fermions, electronic states with non-abelian statistics that can be used to realize topologically protected quantum information processing \cite{nayak_non-abelian_2008,aasen_milestones_2016}. However, topological superconductivity remains elusive in bulk materials and majority of the research is focused on heterostructures coupling conventional superconductors to topological insulators (TIs) \cite{wiedenmann_$4pi$-periodic_2016,bocquillon_gapless_2017, deacon_josephson_2017} or semiconducting structures with strong spin-orbit coupling in the presence of an external magnetic field \cite{Deng2016,Zhang2018}.

One of the first and most important challenges to realize a topologically protected  qubit, is the unambiguous detection of Majorana modes. Recent experimental and theoretical works~\cite{Liu2017,yu_non-majorana_2020} have shown that in general it is very challenging to unambiguously attribute features in d.c. transport measurements to the presence of Majorana modes. These developments show the pressing need to use alternative ways to confidently identify the presence of Majorana states and therefore of topological superconductivity.

Josephson junctions (JJs) have been proposed as a suitable platform to observe localized Majorana modes first in the context of JJs fabricated on 3D TIs \cite{fu_superconducting_2008}, 2D TIs \cite{fu_josephson_2009} and more recently on more conventional III-V heterostructures \cite{hell_two-dimensional_2017, pientka_topological_2017}. JJs are attractive since the phase across the junction provides an additional knob to control the topological phase \cite{fu_josephson_2009, hell_two-dimensional_2017, pientka_topological_2017}. Among the predicted properties of JJs realized on such materials, the expected 4$\pi$ periodicity of the current phase relationship (CPR) has received considerable experimental attention. The 4$\pi$ periodicity of such JJs emerges from the time reversal protected crossing of Andreev bound states (ABS) at a phase of $\pi$ \cite{fu_josephson_2009}, as a consequence it cannot be detected through d.c. measurement since the lower energy state (at zero phase) is not the ground state at 2$\pi$ and the system can thermalize through multiple channels \cite{kwon_fractional_2004, badiane_ac_2013,pikulin_phenomenology_2012}. To circumvent this issue, a.c. driven JJs should provide several features that can be associated to topological superconductivity \cite{wiedenmann_$4pi$-periodic_2016, bocquillon_gapless_2017, rokhinson_fractional_2012, deacon_josephson_2017, laroche_observation_2019}. The preeminent feature is the absence of odd Shapiro steps. However, it was pointed out earlier that Landau-Zener transitions (LZT) between Andreev bound states would lead to the same feature. So far, no experiments have shown that in realistic setups this is the case and all the experiments in which missing Shapiro steps have been observed are consistent with the presence of a topological superconducting state.

A microwave drive, biasing the junction, imposes a periodic modulation of the bias current across the junction which leads to a phase advance of an integer multiple of current phase relationship (CPR) period per cycle. Since the time derivative of the phase is directly proportional to the voltage this leads to constant voltage steps, known as Shapiro steps, in the voltage-current (VI) characteristic of the junction \cite{shapiro_josephson_1963}. In conventional JJs, the CPR is 2$\pi$ periodic \cite{waldram_josephson_1976, kautz_noise_1996} however there has been recent reports of 4$\pi$ periodic CPR which is accompanied by observing missing odd steps in V-I curves. Interestingly, this signature can persist at low frequency even if the CPR has both a 4$\pi$ and a 2$\pi$ periodic component \cite{dominguez_josephson_2017, pico-cortes_signatures_2017}. 

%, if the current bias is large enough

Missing Shapiro steps have been observed in a number of materials systems such as 2D TI HgTe quantum wells (QW) \cite{bocquillon_gapless_2017}, 3D TI HgTe QW \cite{wiedenmann_$4pi$-periodic_2016} , Dirac semimetals \cite{li_4-periodic_2018} and semiconductor nanowires \cite{rokhinson_fractional_2012}. In all these studies, the missing Shapiro steps have been ascribed to the existence of Majorana modes rather than to LZTs between the low energy and the high energy state of an ABS doublet. This has been justified by examining the energy splitting $\Delta E$ between the ABS states which are directly related to the transparency ($\tau$) of the JJ $\Delta E = 2 |\Delta| \sqrt{1 - \tau}$ \cite{beenakker_josephson_1991}. A fast drive can induce transitions from the low energy to the high energy branch and mimic a 4$\pi$ periodic system. However in order to observe well quantized steps, the probability for occurrence of a LZT needs to be near unity as shown in \cite{wiedenmann_$4pi$-periodic_2016} and Supplementary Information (SI), which requires quite stringent high transparency of the JJ as we will discuss.

In this work, we present results obtained on InAs surface QWs coupled to epitaxial Al contacts. In absence of magnetic field, the system is topologically trivial. Still, we observe a missing first Shapiro step and a well quantized second step similar to what has been observed in previous studies on topological insulators and InSb in presence of magnetic field \cite{wiedenmann_$4pi$-periodic_2016, bocquillon_gapless_2017, rokhinson_fractional_2012}. Furthermore, at higher frequency, we observe small half-integer steps consistent with the existence of higher frequency components in the CPR as expected in high transparency JJ. Our results are validated by simulations including LZTs.

\begin{figure}[ht!]
\centering
\includegraphics[width=0.48\textwidth]{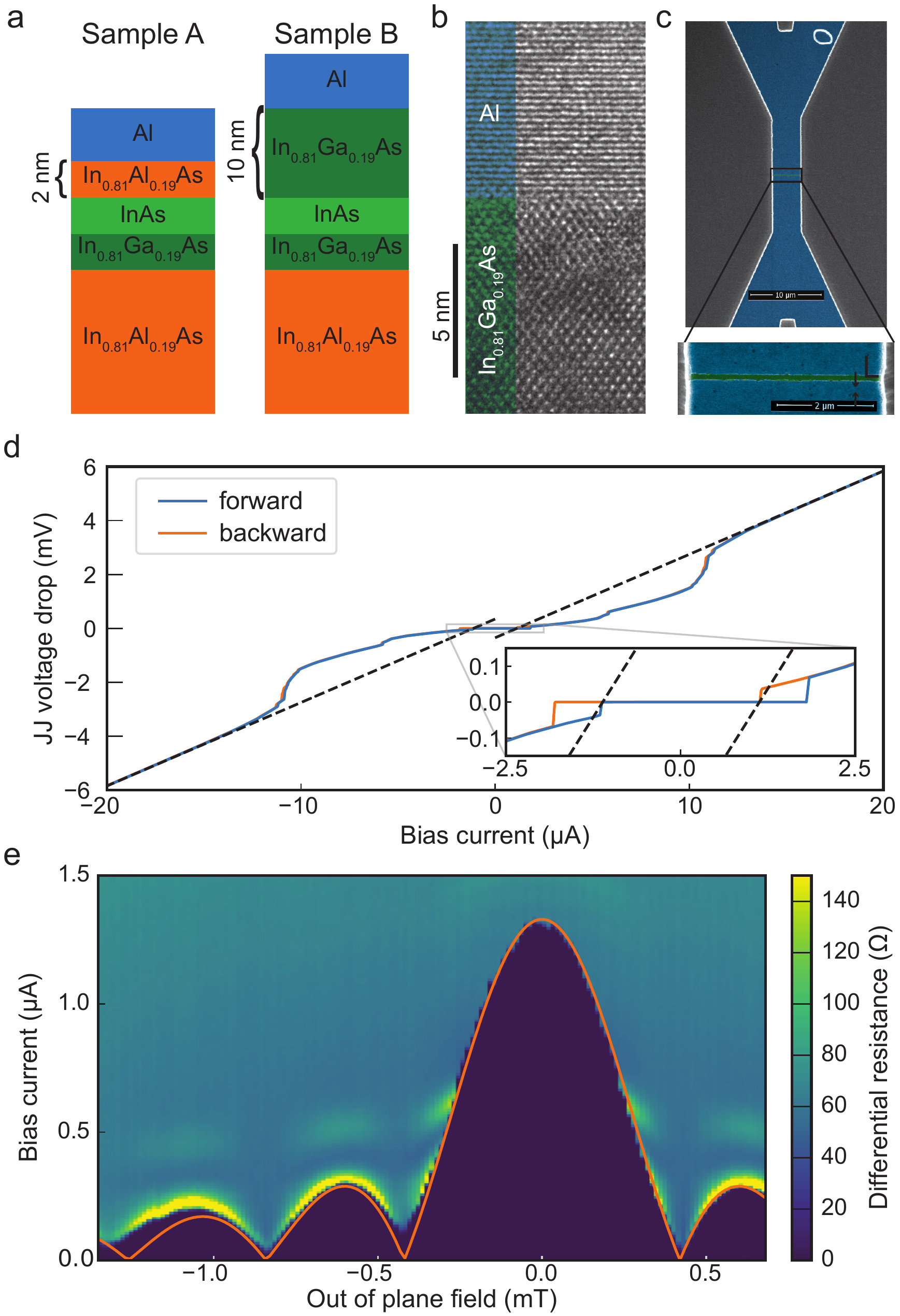}
\caption{\label{figure1}\textbf{a} Sample stack structure. The quantum well consists in a 4nm layer of InAs grown on top of an In$_{0.81}$Ga$_{0.19}$As layer and capped with 2 nm of In$_{0.81}$Al$_{0.19}$As for sample A and 10 nm of In$_{0.81}$Ga$_{0.19}$As for sample B. \textbf{b} TEM picture illustrating the expitaxial nature of the Al-InAs interface. \textbf{c} False-Color SEM image of the Josephson junction.\textbf{d} Voltage-current characteristic of sample A in the absence of microwave irradiation. The dashed line corresponds to linear fit at high bias used to extract the normal resistance and the excess current. \textbf{e} Differential resistance of device A as a function of the bias current and an out-of-plane magnetic field. The overlayed orange curve is the theoretical dependence of $I_c$ for a uniform current distribution.
}
\end{figure}

% introduce material and device fabrication
% introduce both sample and their characteristic Ic, Iexe and Rn. Compute the average transparency based on theory
% mention normal Fraunhofer and homogeneous current distribution
% Insist on the trivial nature (cite Charlie and SQUID paper to introduce the possibility of topology)

This study focuses on two JJs (A and B) fabricated on two slightly different InAs surface QWs with epitaxial Al contacts. The samples are grown on semi-insulating InP (100) substrates \cite{shabani_two-dimensional_2016, wickramasinghe_transport_2018, mayer_superconducting_2019}. The QW consists of a 4 nm layer of InAs grown on a layer of In$_{0.81}$Ga$_{0.19}$ as depicted in Fig.\ref{figure1}a. For sample A, the capping layer is 2 nm of In$_{0.81}$Al$_{0.19}$As , while for sample B, the capping layer is 10 nm of In$_{0.81}$Ga$_{0.19}$As. Fig.\ref{figure1}b is a transmission electron microscope image of the interface between the semiconductor and the Al layer which shows an impurity free interface. These epitaxial interfaces are now widely used in quantum devices to study mesoscopic superconductivity \cite{bottcher_superconducting_2018, mayer_gate_2020}, topological superconductivity \cite{fornieri_evidence_2019, mayer_superconducting_2019} and to develop tunable qubits for quantum information technology \cite{casparis_voltage-controlled_2019}. The gap L between the superconducting contact is 80 nm for device A and 120 nm for device B as illustrated Fig\ref{figure1}c. Details of fabrication and measurements are described in SI.

Figure \ref{figure1}d presents the VI characteristic of sample A in the absence of microwave excitation. The junction is markedly hysteretic but as shown in a previous study \cite{mayer_superconducting_2019} this hysteresis can be ascribed to thermal effects \cite{courtois_origin_2008} rather than capacitive effects. By fitting the linear high current part of the characteristic, we can extract the JJs normal resistance (R$_\textrm{n}$) and the excess current (I$_{\textrm{ex}}$) defined by the intersection of the fit with the x-axis. We report in table \ref{samples} the critical current (I$_\textrm{c}$), along with I$_{\textrm{ex}}$, R$_n$, and the estimated capacitance C from a simple coplanar model \cite{gevorgian_line_2001}. From the excess current we can estimate the average semiconductor-superconductor interface transparency of the junction estimated using the Octavio–Tinkham–Blonder–Klapwijk theory \cite{niebler_analytical_2009}. We obtain 0.86 for Sample A and 0.78 for sample B. The induced superconducting gap is taken to be the one of the Al layer, which we estimate to be 220 $\mu$eV for both samples based on the critical temperature of the film. We report the values of the critical current for both the cold and hot electrons branches where the cold branch goes from 0 bias to high bias and corresponds to a lower effective electronic temperature before the transition out of the superconducting state. The mean free path $l_e$ and the density have been measured on different pieces in a Van der Pauw geometry for both samples. The density is in the range of $1 \times 10^{12}$ cm$^{-2}$ and $l_e$ is about 200 nm, meaning both junctions are nearly ballistic, $L<l_e$, and in the short-junction regime $L\ll\sqrt{l_e\,\xi}$,
with $\xi=\hbar v_F/\pi\Delta\sim 500$~nm the superconducting coherence length estimated
%($E_{\textrm{Thouless}} = \frac{\hbar\,v_F\,l_e}{2\,L^2} \sim 3 meV \ll \Delta$, 
using the lowest density 
%and L from sample B along with 
and an effective mass of 0.04 \cite{yuan_experimental_2019}.

\begin{table}[ht!]
    \centering
    \begin{tabular}{|c|c|c|c|c|c|c|}
        \hline
         & I$_\textrm{c}^{\textrm{cold}}$ ($\mu$A)& I$_c^{\textrm{hot}}$ ($\mu$A)& I$_{\textrm{ex}}$ ($\mu$A) & R$_\textrm{n}$ ($\Omega$) & C (fF)\\
         \hline
         JJ A & 1.8 & 1.1 & 1.1 & 310 & $\sim 1$ \\
         \hline
         JJ B & 5.0 & 4.1 & 3.5 & 97 & $\sim 1$ \\
         \hline
    \end{tabular}
    \caption{Parameters of Sample A and Sample B}
    \label{samples}
\end{table}{}

Figure \ref{figure1}e is a map of the differential resistance of sample A as a function of the bias current and an out of plane magnetic field. The observed Fraunhofer pattern has the expected ratio between the central lobe and the first lobe suggesting a uniform current distribution. The period corresponds to a flux focusing yielding an enhancement of the field by about 3, similar to values reported in \cite{suominen_anomalous_2017}.

\begin{figure*}[ht!]
\centering
\includegraphics[width=0.98\textwidth]{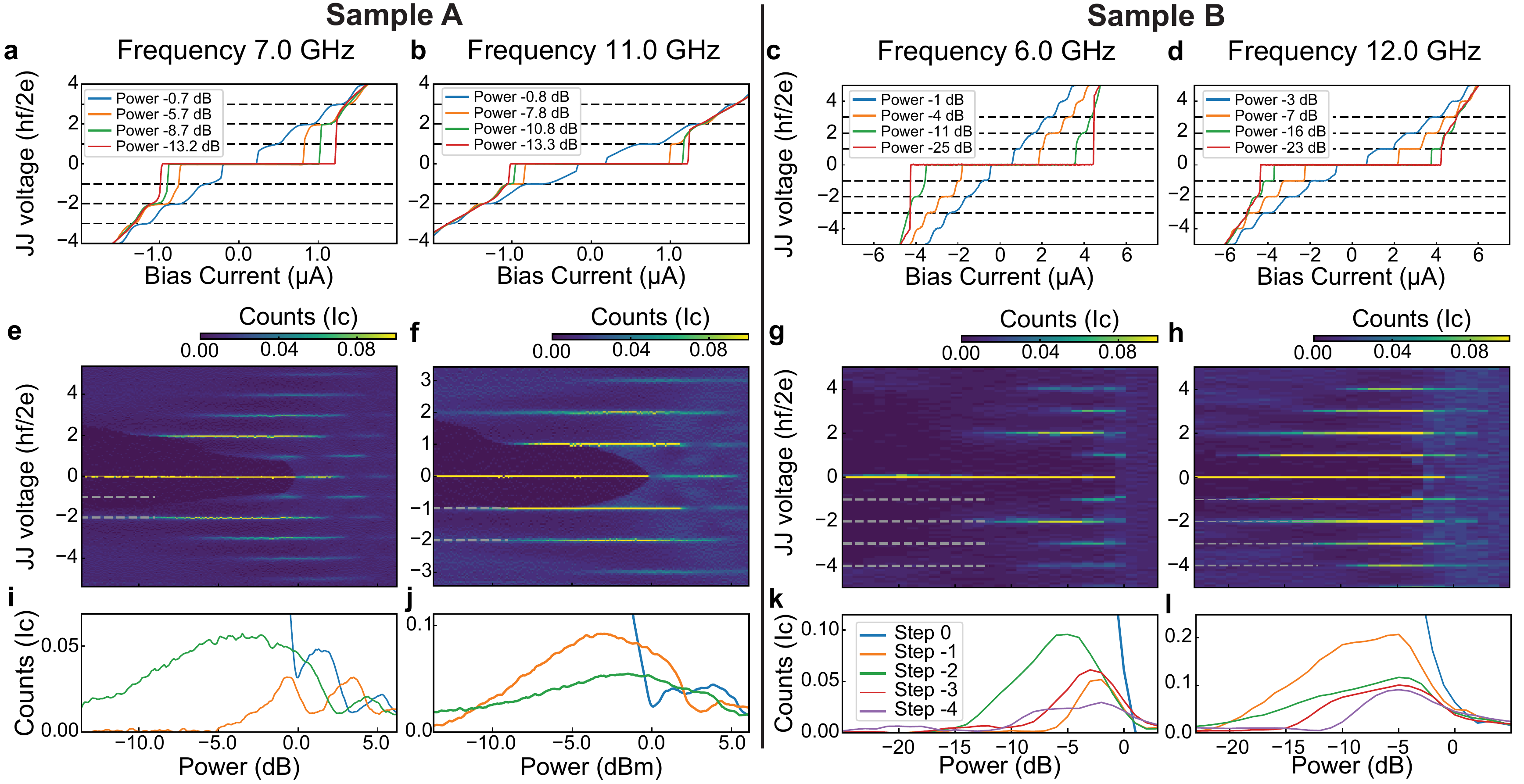}
\caption{\label{figure2}Voltage-current characteristic in the presence of a microwave radiation (a-d), histogram of the Josephson junction voltage as a function of the microwave power (e-h) and width of the Shapiro steps as a function of the microwave power (i-h) for samples A and B at two different frequencies for each sample. In both the histogram and the plot of the width of each steps, counts are expressed in unit of the critical current of the junction in the absence of the microwave drive. In both samples, while at high frequencies all Shapiro steps are visible, at low frequency and low power the first step appears to be strongly suppressed. The microwave power is normalized to the power required to observe the vanishing of the critical current.
}
\end{figure*}

% discuss first line of the figure comparing both sample and frequencies
% Move on to histograms. Point to the weak 3rd step on Sample A
% Discuss limit frequency and number of modes involved in the 4π current

Figure~\ref{figure2} presents results obtained on both sample A and B at two different frequencies of microwave excitation. At high frequency (namely 11 GHz for sample A and 12 GHz for sample B), all the expected steps are visible as can be seen from the VI characteristics presented in Fig.~\ref{figure2} b and d. However at lower frequency (7 GHz for sample A and 6 GHz for sample B), the first Shapiro step is missing in both samples at low power. This fact is particularly clear in the VI characteristics with a microwave power of -5.7 dB (Fig.~\ref{figure2}a) and -4 dB (Fig.~\ref{figure2}c) respectively. On the cold electron branch, the hysteretic behavior of our junctions could lead to missing Shapiro steps since the critical current governing the switching out of the superconducting state is associated with cold electrons and the one governing the dynamics of the steps is associated with hot electrons. However, in the data presented, the first step is missing on both the cold and the hot electron branch and cannot be simply explained by the hysteretic behavior of the junction. In the following we focus our discussion on the hot electron branch of the data.

In order to present the power dependence of the Shapiro steps, we have plotted, for each sample and frequency, the histogram of the voltage distribution as a function of the microwave power in Fig.\ref{figure2} e-h. Finally, in Fig.\ref{figure2} i-l, we plot the weights of the steps as a function of power. As seen already in the VI characteristics, for each sample the first step is strongly suppressed at low frequency and at low power and only emerges close to the drive power required to suppress the 0 volt step. In addition, in the case of sample B, a weak suppression of the third Shapiro step is visible in Fig.~\ref{figure2} g and k.

These experimental observation of missing Shapiro steps are quite similar to the ones obtained on platforms expected to host Majorana modes but in our case, the system is trivial in absence of magentic field. We note that similar InAs QW may host a topological phase \cite{fornieri_evidence_2019, mayer_superconducting_2019}, in presence of a sizable in-plane magnetic field. Data presented here are all taken at zero magnetic field and hence our JJs are topologically trivial. %Therefore, we focus on high transparency modes of our JJs and the possibility of Landau-Zener assisted 4$\pi$ periodicity. 

A microscopic analysis of the dynamics of the JJ taking into account the presence of all the transverse modes, and a biasing current with both a d.c. and a.c. term is computationally prohibitive~\cite{Rossignol2019}. For this reason we describe the JJ via an effective shunted junction model. In general the model has both a resisistive and a capacitive channel. The junctions considered, given their geometry, have a very small capacitance  $C \sim 1$ fF. We therefore neglect the capacitive channel and model the JJ's dynamics using a resistively shunted junction (RSJ) model:
%Let $I_R$ be the current in the resistive channel, and $I_{\rm drive}=I_{\rm dc}+I_{\rm ac}\sin(\omega_{\rm ac}t)$ the external current driving the junction. %Generally, there also exists a capacitive channel but estimating the geometric capacitance of the junction to be $\sim 1$ fF, we neglect the current contribution from the capacitive channel. Using the relation $\dot\phi=(2e/\hbar)V$, from the RSJ model we obtain the following dynamical equation:
%
\begin{equation}
     \frac{\hbar}{2 e R_n} \frac{d\phi}{dt}= I_{dc} + I_{ac} \sin(2\pi f_{ac}t) - I_s(\phi).
    \label{EqMotion}
\end{equation}

where $R_n$ is the normal-state resistance of the junction, $I_{dc}$ is the d.c. bias current, $I_{ac}$ the amplitude of the a.c. current due to the microwave radiation with frequency $f_{ac}$, and $I_s(\phi)$ is the supercurrent. 

To model the supercurrent flowing across the JJ we use two effective modes: a very low transparency mode with a purely sinusoidal CPR and completely negligible probability to undergo a LZT, and an effective mode with very high transparency $\tau$ so that:
\begin{equation}
    I_s(\phi) = I_c \left(s\frac{1-n}{\alpha_{\tau}} \frac{\sin(\phi)}{\sqrt{1 - \tau \sin^2(\phi/2)}} + \frac{n}{\alpha_{0}} \sin(\phi)\right)
    \label{eq:Is}
\end{equation}
where $I_c$ is the experimentally measured critical current, $s=\pm 1$ encodes the switching due to an LZT, 
$n$ is the fraction of the current from the purely sinusoidal mode, 
and $\alpha_{\tau}$ and $\alpha_{0}$ are the values of $\sin(\phi)/\sqrt{1-T \sin^2(\phi/2)}$ and $\sin(\phi)$, respectively,
for the angle $\phi$ for which the current is maximum.

%For the supercurrent $I_s$, we will consider a combination of an {\em effective} high-transparency mode with effective transparency $\tau$ and an {\em effective} low-transparency mode, in analogy with previous analyses of JJ dynamics in the presence of $2\pi$- and $4\pi$-periodic supercurrents \cite{dominguez_josephson_2017, pico-cortes_signatures_2017}. %Perhaps we justify bimodal disrtribution model here 
%For such a high-transparency mode, the Andreev bound state has energy  $E(\phi) = \pm \Delta \sqrt{1 -\tau \sin^2(\phi/2)}$. We assume that the high-transparency effective mode is allowed to participate in a Landau-Zener process at the avoided crossing, resulting in a change in the sign of its supercurrent contribution:
%
%where $s=\pm 1$ encodes the switching due to an LZT, $n$ describes the fraction of the critical current coming from the {\em LZT-inactive} contribution to the supercurrent, and $\alpha_{\tau}$ and $\alpha_{0}$ are normalization parameters to cancel out the amplitude of $\sin(\phi)/\sqrt{1-T \sin^2(\phi/2)}$ and $\sin(\phi)$, respectively, evaluated at the maximum of $I_s$.
For the mode with finite transparency $\tau$ the probability of an LZT is given by \cite{averin_bardas_1995},
\begin{equation}
 P_{LZT}(t) = \exp\left( - \pi\,\frac{\Delta (1-\tau)}{e\,\vert V(t)\vert} \right)
\end{equation}
%
%\noteER{Maybe as a possilbe justification:}
We can attribute the inability of the purely sinusoidal mode to undergo LZTs either to a very small value of the transparency or to the combined effects of disorder and possible phase fluctuations of the order parameter along the transverse direction. Such fluctuations cannot be excluded considering that the width of the junctions is much larger than the superconducting coherence length.
We neglect interference effects due to phase fluctuations and coherence between LZT's. We also do not consider relaxation from higher to lower energy states. %Notice that, via the dependence on $\dot\phi(t)$, $P_{LZT}$ is time dependent. Following Ref. \cite{dominguez_dynamical_2012}, 
By solving Eq.~\ref{EqMotion} accounting for the dynamics due to the LZTs of the effective high-transparency mode, we obtain the time evolution of $\phi$,
and then of the d.c. voltage by time averaging  $V(t)=(\hbar/2e)\dot\phi$. 

\begin{figure}[ht!]
    \centering
    \includegraphics[width=0.48\textwidth]{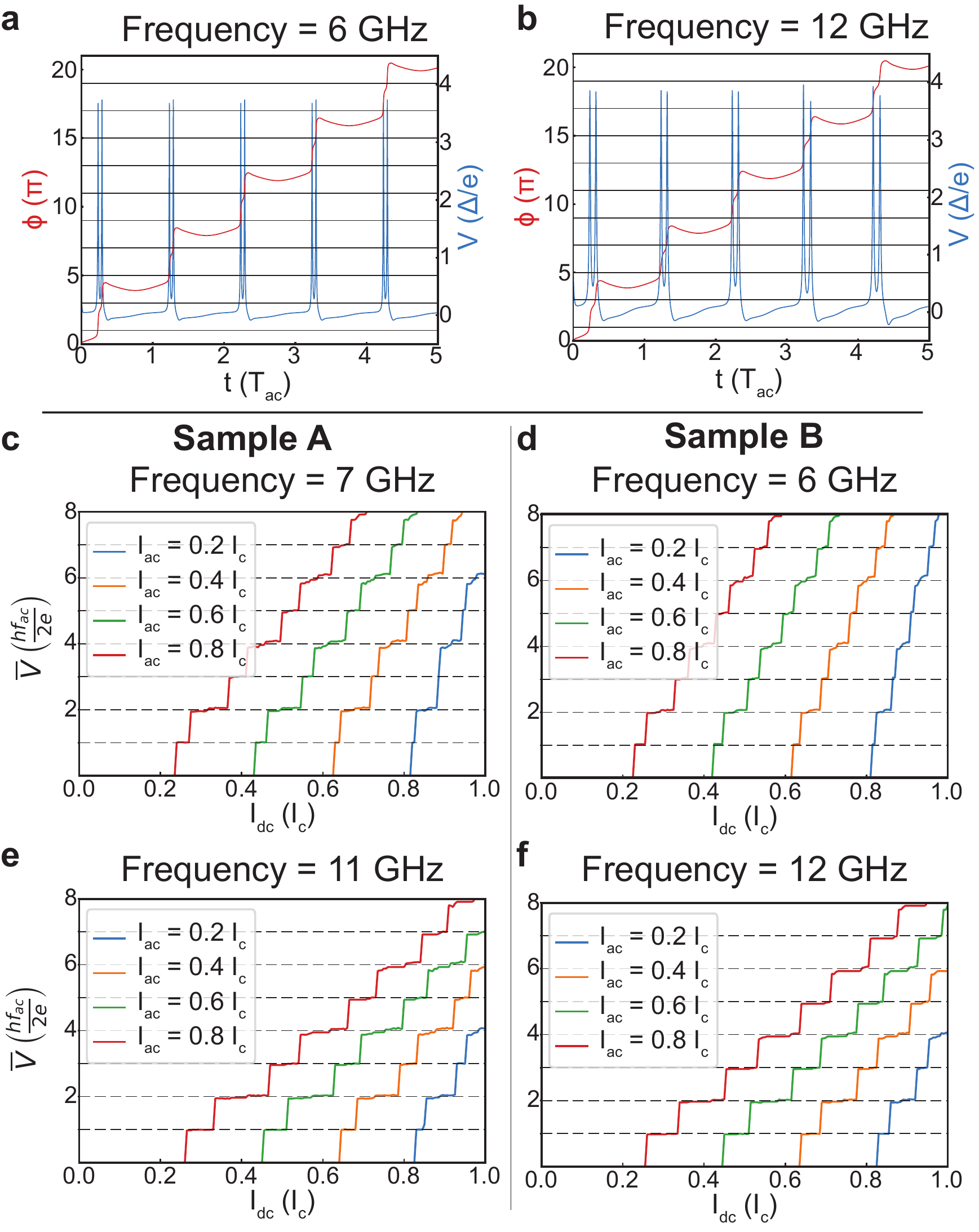}
    \caption{RSJ simulation accounting for LZTs. a, b: Phase and instantaneous voltage across the JJ. The parameters of sample B are used with $\tau = 0.98$ and $n = 0.97$. The driving current are $I_{dc} = 0.7\,I_c$ at 6 GHz and $0.8\,I_c$ at 12 GHz, and $I_{ac} = 0.4\,I_c$ at both frequencies, which corresponds to the second Shapiro step. c, e $I_{dc}$-$\overline{V}$ curves showing Shapiro steps at 7 and 11 GHz for the parameters of sample A, with $\tau = 0.98$ and $n = 0.95$. d, f $I_{dc}$-$\overline{V}$ curves showing Shapiro steps at 6 and 12 GHz for the parameters of sample B, with $\tau = 0.98$ and $n = 0.97$.}
    \label{fig:rcjs}
\end{figure}

Figure~\ref{fig:rcjs} a and b show the dynamic of the phase and the instantaneous voltage in the second Shapiro step using the parameters of sample A. Even though there is a factor of 2 between the drive frequency employed in each case, the value of the instantaneous voltage V(t) at $\pi$ is the same in both cases, as it is dependent on the $I_c\,R_n$ product. This conclusion can be recovered by considering Eq.~\ref{EqMotion} and the fact that inside a step the dynamic of the JJ is phase-locked to the a.c. drive. This implies that the value of the phase maximizing the right hand side of Eq.~\ref{EqMotion} will always occur at the same fraction of the period, leading to a value of $I_{ac} \sin(2\pi f_{ac}t)$ independent of the driving frequency.
This result is important since the lower frequency at which missing steps are observed has been used to estimate the required transparency for LZT to explain missing steps \cite{wiedenmann_$4pi$-periodic_2016}. In the presence of LZT, Shapiro steps are well quantized only if the probability of the transition is very close to 0 or 1 as shown in \cite{wiedenmann_$4pi$-periodic_2016} and SI. In both samples, the induced superconducting gap is close to the bulk gap of the Al layer $\Delta = 220 \mu eV = 53 GHz$, given the high interface transparency, and the lowest frequency at which we observe Shapiro steps is 4 GHz. Using the value of the voltage corresponding to the lowest frequency for which the first odd Shapiro step is missing we would get the transparency of the mode undergoing LZTs should be equal or larger than 0.9985. However, using the value of the voltage corresponding to the {\em frequeny independent} peak of $\dot\phi$ shown in Fig.~\ref{fig:rcjs}, we obtain that  a transparency of 0.98 is sufficient. 

Another consequence of the independence of value of the peak of $\dot{\phi}$ on the drive frequency is that the contribution of LZTs to the 4$\pi$ component in the CPR is also frequency independent.
%and previous results obtained with a constant amount of $4\,\pi$ current can be applied here. 
In the presence of both 2$\pi$ and 4$\pi$ periodic component in the CPR, odd steps are expected to be suppressed when the ac drive frequency is lower than $f_{4\,\pi} = 2\,e\,R_n\,I_{4\,\pi}/h$ \cite{dominguez_josephson_2017, pico-cortes_signatures_2017}. For both our samples, $f_{4\,\pi}$ can be estimated to be about 10 GHz, yielding $I_{4\,\pi} \sim$ 70 nA for sample A and 260 nA for sample B, which in both samples corresponds to about 6\% of the critical current on the hot branch. Using the Al gap value for the induced gap, we can compute, in the short junction limit \cite{beenakker_josephson_1991}, the amount of current carried by a single mode $I_{mode} = \frac{e\,\delta}{2\,\hbar} \sim$ 25 nA. This means that the total 4$\pi$ periodic contribution to the CPR can be assigned to 3(10) modes for the sample A(B). The typical densities reported for each sample yield between 320 and 550 transverse modes in each JJ. This means that only a small minority of the modes (0.5-1\% in sample A and 2-3\% in sample B) need to have near unity transparency and participate LZT processes to explain our observations. Given the small number of modes carrying 4$\pi$ periodic current and the average transparency of the junction, one might expect modes at intermediate transparencies but we do not observe their contributions to LZT in experiment. It remains a theoretical question to why these modes do not participate in LZT processes.% but preserve unity transparency mode contributions.

Figure~\ref{fig:rcjs} c-f presents simulation results for both sample A and B for the same frequencies presented in Fig.~\ref{figure2}. We qualitatively reproduce the experimental results using a slightly smaller I$_{4\,\pi}$ than estimated from the frequency dependence. Additional simulation results are presented in SI.

\begin{figure}[ht!]
\centering
\includegraphics[width=0.48\textwidth]{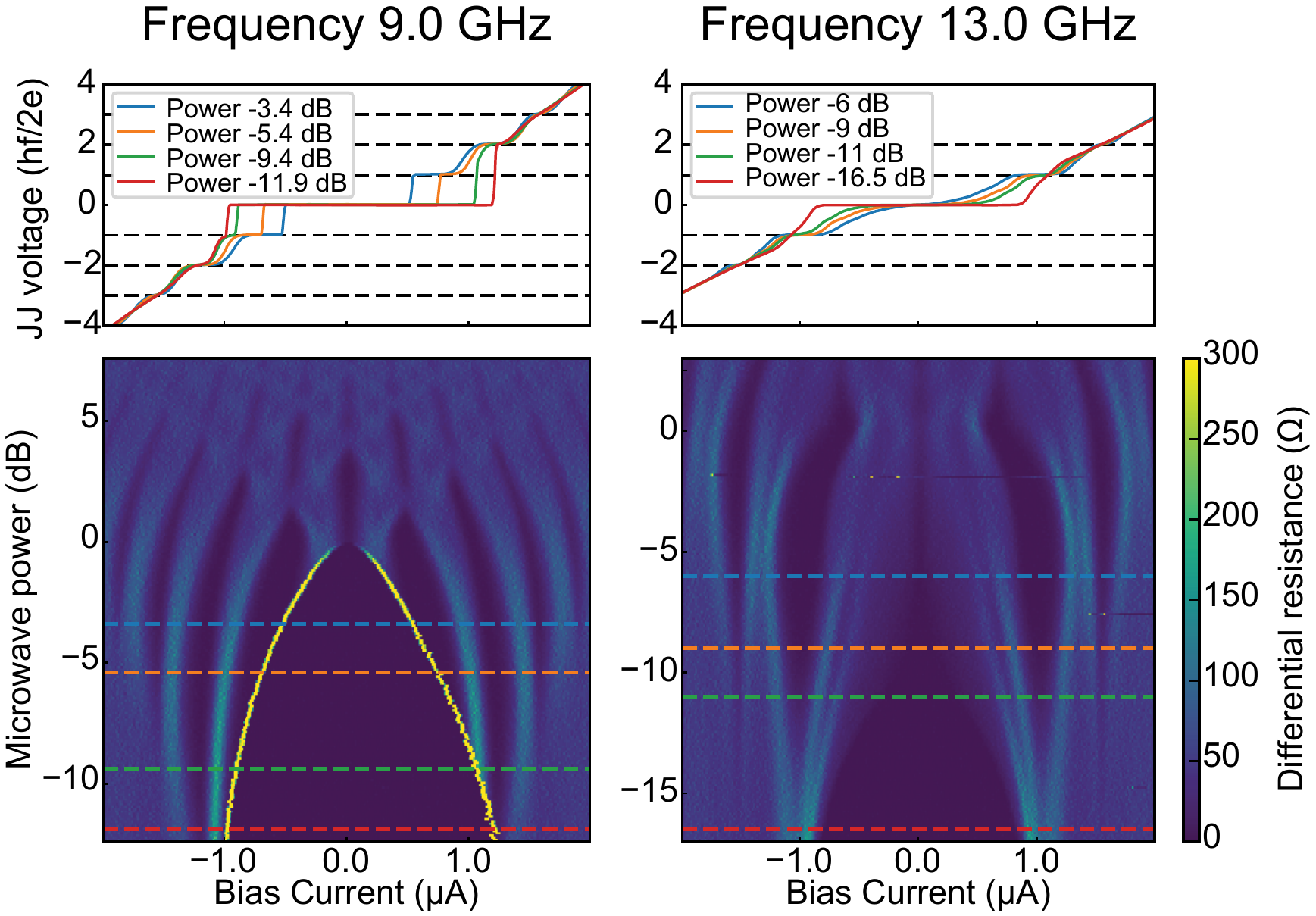}
\caption{\label{figure3}Voltage-current characteristic in the presence of a microwave radiation of sample A and differential resistance map obtained by numerical differentiation as a function of the current bias and applied microwave power. At high frequency, weak half Shapiro steps appear in the VI characteristic which are visible in the differential resistance map as a splitting of finite resistance region separating two integer steps.
}
\end{figure}

% Discuss transparency signatures in CPR and cite SQUID paper, discuss transparency in Shapiro and cite Pangothra 2020

% Describe figure 3 in the context of half integers steps

A JJ with a high transparency is expected to have a forward skewed CPR as can be seen from Eq.~\ref{eq:Is}. The CPR can be directly measured by embedding the JJ in a superconducting interference device (SQUID), whose second JJ has a much larger critical current. Such measurements have been carried out on graphene-based JJ \cite{english_observation_2016}, TI based JJ \cite{sochnikov_nonsinusoidal_2015,kayyalha_highly_2020}, InAs nanowire \cite{spanton_currentphase_2017} and in our samples \cite{mayer_gate_2020}. 
% Use askerzade review as starting point and make the dicsussion quite short
The anharmonicity associated with forward skewness of the CPR has been predicted to lead to the appearance of subharmonics Shapiro steps at high frequency drives\cite{askerzade_effects_2015}. Such subharmonic steps have been observed in several systems \cite{wiedenmann_$4pi$-periodic_2016, snyder_weak-link_2018, lee_ultimately_2015, panghotra_giant_2020}, in which they were associated with a skewed CPR. In Fig.~\ref{figure3}, we present the VI characteristic and differential resistance as a function of a.c. power and bias current of sample A at 9 GHz and 13 GHz. While at 9 GHz, only integer steps are visible, at 13 GHz a weak half-step is visible in the VI. The existence of the sub harmonic step is also visible in the differential resistance map as a splitting of the peak in resistance between the integer steps. This signature provides an additional experimental signature of the high transparency of our junctions.\\

%{Conclusions.} 

In this work we show experimentally that in JJs that are undoubtedly in a topologically trivial phase, for microwave powers and frequencies reported, there are missing odd Shapiro steps consistent with the 4$\pi$ periodic current-phase relation of a topological JJ. We attribute our measurement to the very high transparency of a fraction of the modes in our JJ combined with large value of $I_c R_n$. Our results clearly show that caution should be used to attribute missing Shapiro steps to the presence of Majorana modes. They provide essential guidance to future experiments to use JJs to unambiguously establish the presence of topological superconductivity, and, more in general, significantly enhance our understanding of high quality JJs.

NYU team is supported by NSF DMR Grant No. 1836687 and DARPA Grant No. DP18AP900007. Joseph Yuan acknowledges funding from the ARO/LPS QuaCGR fellowship. JJC and ER
acknowledge support from ARO Grant No. W911NF-18-1-0290 and NSF CAREER Grant No. DMR-1455233.

%\bibliography{biblio}

\begin{thebibliography}{0}
\expandafter\ifx\csname natexlab\endcsname\relax\def\natexlab#1{#1}\fi
\expandafter\ifx\csname bibnamefont\endcsname\relax
  \def\bibnamefont#1{#1}\fi
\expandafter\ifx\csname bibfnamefont\endcsname\relax
  \def\bibfnamefont#1{#1}\fi
\expandafter\ifx\csname citenamefont\endcsname\relax
  \def\citenamefont#1{#1}\fi
\expandafter\ifx\csname url\endcsname\relax
  \def\url#1{\texttt{#1}}\fi
\expandafter\ifx\csname urlprefix\endcsname\relax\def\urlprefix{URL }\fi
\providecommand{\bibinfo}[2]{#2}
\providecommand{\eprint}[2][]{\url{#2}}

\end{thebibliography}


\begin{thebibliography}{50}
\expandafter\ifx\csname natexlab\endcsname\relax\def\natexlab#1{#1}\fi
\expandafter\ifx\csname bibnamefont\endcsname\relax
  \def\bibnamefont#1{#1}\fi
\expandafter\ifx\csname bibfnamefont\endcsname\relax
  \def\bibfnamefont#1{#1}\fi
\expandafter\ifx\csname citenamefont\endcsname\relax
  \def\citenamefont#1{#1}\fi
\expandafter\ifx\csname url\endcsname\relax
  \def\url#1{\texttt{#1}}\fi
\expandafter\ifx\csname urlprefix\endcsname\relax\def\urlprefix{URL }\fi
\providecommand{\bibinfo}[2]{#2}
\providecommand{\eprint}[2][]{\url{#2}}

\bibitem[{\citenamefont{Cao et~al.}(2018)\citenamefont{Cao, Fatemi, Fang,
  Watanabe, Taniguchi, Kaxiras, and Jarillo-Herrero}}]{Cao_Gr_2018}
\bibinfo{author}{\bibfnamefont{Y.}~\bibnamefont{Cao}},
  \bibinfo{author}{\bibfnamefont{V.}~\bibnamefont{Fatemi}},
  \bibinfo{author}{\bibfnamefont{S.}~\bibnamefont{Fang}},
  \bibinfo{author}{\bibfnamefont{K.}~\bibnamefont{Watanabe}},
  \bibinfo{author}{\bibfnamefont{T.}~\bibnamefont{Taniguchi}},
  \bibinfo{author}{\bibfnamefont{E.}~\bibnamefont{Kaxiras}}, \bibnamefont{and}
  \bibinfo{author}{\bibfnamefont{P.}~\bibnamefont{Jarillo-Herrero}},
  \bibinfo{journal}{Nature} \textbf{\bibinfo{volume}{556}}, \bibinfo{pages}{43}
  (\bibinfo{year}{2018}).

\bibitem[{\citenamefont{Fu and Kane}(2008)}]{fu_superconducting_2008}
\bibinfo{author}{\bibfnamefont{L.}~\bibnamefont{Fu}} \bibnamefont{and}
  \bibinfo{author}{\bibfnamefont{C.~L.} \bibnamefont{Kane}},
  \bibinfo{journal}{Phys. Rev. Lett.} \textbf{\bibinfo{volume}{100}},
  \bibinfo{pages}{096407} (\bibinfo{year}{2008}).

\bibitem[{\citenamefont{Oreg et~al.}(2010)\citenamefont{Oreg, Refael, and von
  Oppen}}]{oreg_helical_2010}
\bibinfo{author}{\bibfnamefont{Y.}~\bibnamefont{Oreg}},
  \bibinfo{author}{\bibfnamefont{G.}~\bibnamefont{Refael}}, \bibnamefont{and}
  \bibinfo{author}{\bibfnamefont{F.}~\bibnamefont{von Oppen}},
  \bibinfo{journal}{Phys. Rev. Lett.} \textbf{\bibinfo{volume}{105}},
  \bibinfo{pages}{177002} (\bibinfo{year}{2010}).

\bibitem[{\citenamefont{Lutchyn et~al.}(2010)\citenamefont{Lutchyn, Sau, and
  Das~Sarma}}]{Lutchyn_MF_2010}
\bibinfo{author}{\bibfnamefont{R.~M.} \bibnamefont{Lutchyn}},
  \bibinfo{author}{\bibfnamefont{J.~D.} \bibnamefont{Sau}}, \bibnamefont{and}
  \bibinfo{author}{\bibfnamefont{S.}~\bibnamefont{Das~Sarma}},
  \bibinfo{journal}{Phys. Rev. Lett.} \textbf{\bibinfo{volume}{105}},
  \bibinfo{pages}{077001} (\bibinfo{year}{2010}).

\bibitem[{\citenamefont{Nayak et~al.}(2008)\citenamefont{Nayak, Simon, Stern,
  Freedman, and Das~Sarma}}]{nayak_non-abelian_2008}
\bibinfo{author}{\bibfnamefont{C.}~\bibnamefont{Nayak}},
  \bibinfo{author}{\bibfnamefont{S.~H.} \bibnamefont{Simon}},
  \bibinfo{author}{\bibfnamefont{A.}~\bibnamefont{Stern}},
  \bibinfo{author}{\bibfnamefont{M.}~\bibnamefont{Freedman}}, \bibnamefont{and}
  \bibinfo{author}{\bibfnamefont{S.}~\bibnamefont{Das~Sarma}},
  \bibinfo{journal}{Rev. Mod. Phys.} \textbf{\bibinfo{volume}{80}},
  \bibinfo{pages}{1083} (\bibinfo{year}{2008}).

\bibitem[{\citenamefont{Aasen et~al.}(2016)\citenamefont{Aasen, Hell, Mishmash,
  Higginbotham, Danon, Leijnse, Jespersen, Folk, Marcus, Flensberg
  et~al.}}]{aasen_milestones_2016}
\bibinfo{author}{\bibfnamefont{D.}~\bibnamefont{Aasen}},
  \bibinfo{author}{\bibfnamefont{M.}~\bibnamefont{Hell}},
  \bibinfo{author}{\bibfnamefont{R.~V.} \bibnamefont{Mishmash}},
  \bibinfo{author}{\bibfnamefont{A.}~\bibnamefont{Higginbotham}},
  \bibinfo{author}{\bibfnamefont{J.}~\bibnamefont{Danon}},
  \bibinfo{author}{\bibfnamefont{M.}~\bibnamefont{Leijnse}},
  \bibinfo{author}{\bibfnamefont{T.~S.} \bibnamefont{Jespersen}},
  \bibinfo{author}{\bibfnamefont{J.~A.} \bibnamefont{Folk}},
  \bibinfo{author}{\bibfnamefont{C.~M.} \bibnamefont{Marcus}},
  \bibinfo{author}{\bibfnamefont{K.}~\bibnamefont{Flensberg}},
  \bibnamefont{et~al.}, \bibinfo{journal}{Phys. Rev. X}
  \textbf{\bibinfo{volume}{6}}, \bibinfo{pages}{031016} (\bibinfo{year}{2016}).

\bibitem[{\citenamefont{Wiedenmann et~al.}(2016)\citenamefont{Wiedenmann,
  Bocquillon, Deacon, Hartinger, Herrmann, Klapwijk, Maier, Ames, Brüne, Gould
  et~al.}}]{wiedenmann_$4pi$-periodic_2016}
\bibinfo{author}{\bibfnamefont{J.}~\bibnamefont{Wiedenmann}},
  \bibinfo{author}{\bibfnamefont{E.}~\bibnamefont{Bocquillon}},
  \bibinfo{author}{\bibfnamefont{R.~S.} \bibnamefont{Deacon}},
  \bibinfo{author}{\bibfnamefont{S.}~\bibnamefont{Hartinger}},
  \bibinfo{author}{\bibfnamefont{O.}~\bibnamefont{Herrmann}},
  \bibinfo{author}{\bibfnamefont{T.~M.} \bibnamefont{Klapwijk}},
  \bibinfo{author}{\bibfnamefont{L.}~\bibnamefont{Maier}},
  \bibinfo{author}{\bibfnamefont{C.}~\bibnamefont{Ames}},
  \bibinfo{author}{\bibfnamefont{C.}~\bibnamefont{Brüne}},
  \bibinfo{author}{\bibfnamefont{C.}~\bibnamefont{Gould}},
  \bibnamefont{et~al.}, \bibinfo{journal}{Nature Communications}
  \textbf{\bibinfo{volume}{7}} (\bibinfo{year}{2016}).

\bibitem[{\citenamefont{Bocquillon et~al.}(2017)\citenamefont{Bocquillon,
  Deacon, Wiedenmann, Leubner, Klapwijk, Brüne, Ishibashi, Buhmann, and
  Molenkamp}}]{bocquillon_gapless_2017}
\bibinfo{author}{\bibfnamefont{E.}~\bibnamefont{Bocquillon}},
  \bibinfo{author}{\bibfnamefont{R.~S.} \bibnamefont{Deacon}},
  \bibinfo{author}{\bibfnamefont{J.}~\bibnamefont{Wiedenmann}},
  \bibinfo{author}{\bibfnamefont{P.}~\bibnamefont{Leubner}},
  \bibinfo{author}{\bibfnamefont{T.~M.} \bibnamefont{Klapwijk}},
  \bibinfo{author}{\bibfnamefont{C.}~\bibnamefont{Brüne}},
  \bibinfo{author}{\bibfnamefont{K.}~\bibnamefont{Ishibashi}},
  \bibinfo{author}{\bibfnamefont{H.}~\bibnamefont{Buhmann}}, \bibnamefont{and}
  \bibinfo{author}{\bibfnamefont{L.~W.} \bibnamefont{Molenkamp}},
  \bibinfo{journal}{Nature Nanotech} \textbf{\bibinfo{volume}{12}},
  \bibinfo{pages}{137} (\bibinfo{year}{2017}).

\bibitem[{\citenamefont{Deacon et~al.}(2017)\citenamefont{Deacon, Wiedenmann,
  Bocquillon, Domínguez, Klapwijk, Leubner, Brüne, Hankiewicz, Tarucha,
  Ishibashi et~al.}}]{deacon_josephson_2017}
\bibinfo{author}{\bibfnamefont{R.}~\bibnamefont{Deacon}},
  \bibinfo{author}{\bibfnamefont{J.}~\bibnamefont{Wiedenmann}},
  \bibinfo{author}{\bibfnamefont{E.}~\bibnamefont{Bocquillon}},
  \bibinfo{author}{\bibfnamefont{F.}~\bibnamefont{Domínguez}},
  \bibinfo{author}{\bibfnamefont{T.}~\bibnamefont{Klapwijk}},
  \bibinfo{author}{\bibfnamefont{P.}~\bibnamefont{Leubner}},
  \bibinfo{author}{\bibfnamefont{C.}~\bibnamefont{Brüne}},
  \bibinfo{author}{\bibfnamefont{E.}~\bibnamefont{Hankiewicz}},
  \bibinfo{author}{\bibfnamefont{S.}~\bibnamefont{Tarucha}},
  \bibinfo{author}{\bibfnamefont{K.}~\bibnamefont{Ishibashi}},
  \bibnamefont{et~al.}, \bibinfo{journal}{Phys. Rev. X}
  \textbf{\bibinfo{volume}{7}}, \bibinfo{pages}{021011} (\bibinfo{year}{2017}).

\bibitem[{\citenamefont{Deng et~al.}(2016)\citenamefont{Deng, Vaitiekenas,
  Hansen, Danon, Leijnse, Flensberg, Nyg{\aa}rd, Krogstrup, and
  Marcus}}]{Deng2016}
\bibinfo{author}{\bibfnamefont{M.~T.} \bibnamefont{Deng}},
  \bibinfo{author}{\bibfnamefont{S.}~\bibnamefont{Vaitiekenas}},
  \bibinfo{author}{\bibfnamefont{E.~B.} \bibnamefont{Hansen}},
  \bibinfo{author}{\bibfnamefont{J.}~\bibnamefont{Danon}},
  \bibinfo{author}{\bibfnamefont{M.}~\bibnamefont{Leijnse}},
  \bibinfo{author}{\bibfnamefont{K.}~\bibnamefont{Flensberg}},
  \bibinfo{author}{\bibfnamefont{J.}~\bibnamefont{Nyg{\aa}rd}},
  \bibinfo{author}{\bibfnamefont{P.}~\bibnamefont{Krogstrup}},
  \bibnamefont{and} \bibinfo{author}{\bibfnamefont{C.~M.}
  \bibnamefont{Marcus}}, \bibinfo{journal}{Science}
  \textbf{\bibinfo{volume}{354}}, \bibinfo{pages}{1557} (\bibinfo{year}{2016}),
  ISSN \bibinfo{issn}{0036-8075}.

\bibitem[{\citenamefont{Zhang et~al.}(2018)\citenamefont{Zhang, Liu,
  Gazibegovic, Xu, Logan, Wang, van Loo, Bommer, de~Moor, Car
  et~al.}}]{Zhang2018}
\bibinfo{author}{\bibfnamefont{H.}~\bibnamefont{Zhang}},
  \bibinfo{author}{\bibfnamefont{C.-X.} \bibnamefont{Liu}},
  \bibinfo{author}{\bibfnamefont{S.}~\bibnamefont{Gazibegovic}},
  \bibinfo{author}{\bibfnamefont{D.}~\bibnamefont{Xu}},
  \bibinfo{author}{\bibfnamefont{J.~A.} \bibnamefont{Logan}},
  \bibinfo{author}{\bibfnamefont{G.}~\bibnamefont{Wang}},
  \bibinfo{author}{\bibfnamefont{N.}~\bibnamefont{van Loo}},
  \bibinfo{author}{\bibfnamefont{J.~D.~S.} \bibnamefont{Bommer}},
  \bibinfo{author}{\bibfnamefont{M.~W.~A.} \bibnamefont{de~Moor}},
  \bibinfo{author}{\bibfnamefont{D.}~\bibnamefont{Car}}, \bibnamefont{et~al.},
  \bibinfo{journal}{Nature} \textbf{\bibinfo{volume}{556}}, \bibinfo{pages}{74}
  (\bibinfo{year}{2018}).

\bibitem[{\citenamefont{Liu et~al.}(2017)\citenamefont{Liu, Sau, Stanescu, and
  Das~Sarma}}]{Liu2017}
\bibinfo{author}{\bibfnamefont{C.-X.} \bibnamefont{Liu}},
  \bibinfo{author}{\bibfnamefont{J.~D.} \bibnamefont{Sau}},
  \bibinfo{author}{\bibfnamefont{T.~D.} \bibnamefont{Stanescu}},
  \bibnamefont{and}
  \bibinfo{author}{\bibfnamefont{S.}~\bibnamefont{Das~Sarma}},
  \bibinfo{journal}{Phys. Rev. B} \textbf{\bibinfo{volume}{96}},
  \bibinfo{pages}{075161} (\bibinfo{year}{2017}).

\bibitem[{\citenamefont{Yu et~al.}(2020)\citenamefont{Yu, Chen, Gomanko,
  Badawy, Bakkers, Zuo, Mourik, and Frolov}}]{yu_non-majorana_2020}
\bibinfo{author}{\bibfnamefont{P.}~\bibnamefont{Yu}},
  \bibinfo{author}{\bibfnamefont{J.}~\bibnamefont{Chen}},
  \bibinfo{author}{\bibfnamefont{M.}~\bibnamefont{Gomanko}},
  \bibinfo{author}{\bibfnamefont{G.}~\bibnamefont{Badawy}},
  \bibinfo{author}{\bibfnamefont{E.~P. A.~M.} \bibnamefont{Bakkers}},
  \bibinfo{author}{\bibfnamefont{K.}~\bibnamefont{Zuo}},
  \bibinfo{author}{\bibfnamefont{V.}~\bibnamefont{Mourik}}, \bibnamefont{and}
  \bibinfo{author}{\bibfnamefont{S.~M.} \bibnamefont{Frolov}},
  \bibinfo{journal}{arXiv:2004.08583 [cond-mat]}  (\bibinfo{year}{2020}),
  \bibinfo{note}{arXiv: 2004.08583}.

\bibitem[{\citenamefont{Fu and Kane}(2009)}]{fu_josephson_2009}
\bibinfo{author}{\bibfnamefont{L.}~\bibnamefont{Fu}} \bibnamefont{and}
  \bibinfo{author}{\bibfnamefont{C.~L.} \bibnamefont{Kane}},
  \bibinfo{journal}{Phys. Rev. B} \textbf{\bibinfo{volume}{79}},
  \bibinfo{pages}{161408} (\bibinfo{year}{2009}).

\bibitem[{\citenamefont{Hell et~al.}(2017)\citenamefont{Hell, Leijnse, and
  Flensberg}}]{hell_two-dimensional_2017}
\bibinfo{author}{\bibfnamefont{M.}~\bibnamefont{Hell}},
  \bibinfo{author}{\bibfnamefont{M.}~\bibnamefont{Leijnse}}, \bibnamefont{and}
  \bibinfo{author}{\bibfnamefont{K.}~\bibnamefont{Flensberg}},
  \bibinfo{journal}{Phys. Rev. Lett.} \textbf{\bibinfo{volume}{118}},
  \bibinfo{pages}{107701} (\bibinfo{year}{2017}).

\bibitem[{\citenamefont{Pientka et~al.}(2017)\citenamefont{Pientka, Keselman,
  Berg, Yacoby, Stern, and Halperin}}]{pientka_topological_2017}
\bibinfo{author}{\bibfnamefont{F.}~\bibnamefont{Pientka}},
  \bibinfo{author}{\bibfnamefont{A.}~\bibnamefont{Keselman}},
  \bibinfo{author}{\bibfnamefont{E.}~\bibnamefont{Berg}},
  \bibinfo{author}{\bibfnamefont{A.}~\bibnamefont{Yacoby}},
  \bibinfo{author}{\bibfnamefont{A.}~\bibnamefont{Stern}}, \bibnamefont{and}
  \bibinfo{author}{\bibfnamefont{B.~I.} \bibnamefont{Halperin}},
  \bibinfo{journal}{Phys. Rev. X} \textbf{\bibinfo{volume}{7}},
  \bibinfo{pages}{021032} (\bibinfo{year}{2017}).

\bibitem[{\citenamefont{Kwon et~al.}(2004)\citenamefont{Kwon, Sengupta, and
  Yakovenko}}]{kwon_fractional_2004}
\bibinfo{author}{\bibfnamefont{H.-J.} \bibnamefont{Kwon}},
  \bibinfo{author}{\bibfnamefont{K.}~\bibnamefont{Sengupta}}, \bibnamefont{and}
  \bibinfo{author}{\bibfnamefont{V.~M.} \bibnamefont{Yakovenko}},
  \bibinfo{journal}{Eur. Phys. J. B} \textbf{\bibinfo{volume}{37}},
  \bibinfo{pages}{349} (\bibinfo{year}{2004}).

\bibitem[{\citenamefont{Badiane et~al.}(2013)\citenamefont{Badiane, Glazman,
  Houzet, and Meyer}}]{badiane_ac_2013}
\bibinfo{author}{\bibfnamefont{D.~M.} \bibnamefont{Badiane}},
  \bibinfo{author}{\bibfnamefont{L.~I.} \bibnamefont{Glazman}},
  \bibinfo{author}{\bibfnamefont{M.}~\bibnamefont{Houzet}}, \bibnamefont{and}
  \bibinfo{author}{\bibfnamefont{J.~S.} \bibnamefont{Meyer}},
  \bibinfo{journal}{Comptes Rendus Physique} \textbf{\bibinfo{volume}{14}},
  \bibinfo{pages}{840} (\bibinfo{year}{2013}).

\bibitem[{\citenamefont{Pikulin and
  Nazarov}(2012)}]{pikulin_phenomenology_2012}
\bibinfo{author}{\bibfnamefont{D.~I.} \bibnamefont{Pikulin}} \bibnamefont{and}
  \bibinfo{author}{\bibfnamefont{Y.~V.} \bibnamefont{Nazarov}},
  \bibinfo{journal}{Phys. Rev. B} \textbf{\bibinfo{volume}{86}},
  \bibinfo{pages}{140504} (\bibinfo{year}{2012}).

\bibitem[{\citenamefont{Rokhinson et~al.}(2012)\citenamefont{Rokhinson, Liu,
  and Furdyna}}]{rokhinson_fractional_2012}
\bibinfo{author}{\bibfnamefont{L.~P.} \bibnamefont{Rokhinson}},
  \bibinfo{author}{\bibfnamefont{X.}~\bibnamefont{Liu}}, \bibnamefont{and}
  \bibinfo{author}{\bibfnamefont{J.~K.} \bibnamefont{Furdyna}},
  \bibinfo{journal}{Nature Physics} \textbf{\bibinfo{volume}{8}},
  \bibinfo{pages}{795} (\bibinfo{year}{2012}).

\bibitem[{\citenamefont{Laroche et~al.}(2019)\citenamefont{Laroche, Bouman,
  Woerkom, Proutski, Murthy, Pikulin, Nayak, Gulik, Nygård, Krogstrup
  et~al.}}]{laroche_observation_2019}
\bibinfo{author}{\bibfnamefont{D.}~\bibnamefont{Laroche}},
  \bibinfo{author}{\bibfnamefont{D.}~\bibnamefont{Bouman}},
  \bibinfo{author}{\bibfnamefont{D.~J.~v.} \bibnamefont{Woerkom}},
  \bibinfo{author}{\bibfnamefont{A.}~\bibnamefont{Proutski}},
  \bibinfo{author}{\bibfnamefont{C.}~\bibnamefont{Murthy}},
  \bibinfo{author}{\bibfnamefont{D.~I.} \bibnamefont{Pikulin}},
  \bibinfo{author}{\bibfnamefont{C.}~\bibnamefont{Nayak}},
  \bibinfo{author}{\bibfnamefont{R.~J. J.~v.} \bibnamefont{Gulik}},
  \bibinfo{author}{\bibfnamefont{J.}~\bibnamefont{Nygård}},
  \bibinfo{author}{\bibfnamefont{P.}~\bibnamefont{Krogstrup}},
  \bibnamefont{et~al.}, \bibinfo{journal}{Nat Commun}
  \textbf{\bibinfo{volume}{10}}, \bibinfo{pages}{1} (\bibinfo{year}{2019}),
  ISSN \bibinfo{issn}{2041-1723}, \bibinfo{note}{number: 1 Publisher: Nature
  Publishing Group}.

\bibitem[{\citenamefont{Shapiro}(1963)}]{shapiro_josephson_1963}
\bibinfo{author}{\bibfnamefont{S.}~\bibnamefont{Shapiro}},
  \bibinfo{journal}{Phys. Rev. Lett.} \textbf{\bibinfo{volume}{11}},
  \bibinfo{pages}{80} (\bibinfo{year}{1963}), \bibinfo{note}{publisher:
  American Physical Society}.

\bibitem[{\citenamefont{Waldram}(1976)}]{waldram_josephson_1976}
\bibinfo{author}{\bibfnamefont{J.~R.} \bibnamefont{Waldram}},
  \bibinfo{journal}{Rep. Prog. Phys.} \textbf{\bibinfo{volume}{39}},
  \bibinfo{pages}{751} (\bibinfo{year}{1976}), ISSN \bibinfo{issn}{0034-4885},
  \bibinfo{note}{publisher: IOP Publishing}.

\bibitem[{\citenamefont{Kautz}(1996)}]{kautz_noise_1996}
\bibinfo{author}{\bibfnamefont{R.~L.} \bibnamefont{Kautz}},
  \bibinfo{journal}{Rep. Prog. Phys.} \textbf{\bibinfo{volume}{59}},
  \bibinfo{pages}{935} (\bibinfo{year}{1996}), ISSN \bibinfo{issn}{0034-4885},
  \bibinfo{note}{publisher: IOP Publishing}.

\bibitem[{\citenamefont{Domínguez et~al.}(2017)\citenamefont{Domínguez,
  Kashuba, Bocquillon, Wiedenmann, Deacon, Klapwijk, Platero, Molenkamp,
  Trauzettel, and Hankiewicz}}]{dominguez_josephson_2017}
\bibinfo{author}{\bibfnamefont{F.}~\bibnamefont{Domínguez}},
  \bibinfo{author}{\bibfnamefont{O.}~\bibnamefont{Kashuba}},
  \bibinfo{author}{\bibfnamefont{E.}~\bibnamefont{Bocquillon}},
  \bibinfo{author}{\bibfnamefont{J.}~\bibnamefont{Wiedenmann}},
  \bibinfo{author}{\bibfnamefont{R.~S.} \bibnamefont{Deacon}},
  \bibinfo{author}{\bibfnamefont{T.~M.} \bibnamefont{Klapwijk}},
  \bibinfo{author}{\bibfnamefont{G.}~\bibnamefont{Platero}},
  \bibinfo{author}{\bibfnamefont{L.~W.} \bibnamefont{Molenkamp}},
  \bibinfo{author}{\bibfnamefont{B.}~\bibnamefont{Trauzettel}},
  \bibnamefont{and} \bibinfo{author}{\bibfnamefont{E.~M.}
  \bibnamefont{Hankiewicz}}, \bibinfo{journal}{Phys. Rev. B}
  \textbf{\bibinfo{volume}{95}}, \bibinfo{pages}{195430}
  (\bibinfo{year}{2017}).

\bibitem[{\citenamefont{Picó-Cortés et~al.}(2017)\citenamefont{Picó-Cortés,
  Domínguez, and Platero}}]{pico-cortes_signatures_2017}
\bibinfo{author}{\bibfnamefont{J.}~\bibnamefont{Picó-Cortés}},
  \bibinfo{author}{\bibfnamefont{F.}~\bibnamefont{Domínguez}},
  \bibnamefont{and} \bibinfo{author}{\bibfnamefont{G.}~\bibnamefont{Platero}},
  \bibinfo{journal}{Phys. Rev. B} \textbf{\bibinfo{volume}{96}},
  \bibinfo{pages}{125438} (\bibinfo{year}{2017}).

\bibitem[{\citenamefont{Li et~al.}(2018)\citenamefont{Li, Boer, Ronde,
  Ramankutty, Heumen, Huang, Visser, Golubov, Golden, and
  Brinkman}}]{li_4-periodic_2018}
\bibinfo{author}{\bibfnamefont{C.}~\bibnamefont{Li}},
  \bibinfo{author}{\bibfnamefont{J.~C.~d.} \bibnamefont{Boer}},
  \bibinfo{author}{\bibfnamefont{B.~d.} \bibnamefont{Ronde}},
  \bibinfo{author}{\bibfnamefont{S.~V.} \bibnamefont{Ramankutty}},
  \bibinfo{author}{\bibfnamefont{E.~v.} \bibnamefont{Heumen}},
  \bibinfo{author}{\bibfnamefont{Y.}~\bibnamefont{Huang}},
  \bibinfo{author}{\bibfnamefont{A.~d.} \bibnamefont{Visser}},
  \bibinfo{author}{\bibfnamefont{A.~A.} \bibnamefont{Golubov}},
  \bibinfo{author}{\bibfnamefont{M.~S.} \bibnamefont{Golden}},
  \bibnamefont{and} \bibinfo{author}{\bibfnamefont{A.}~\bibnamefont{Brinkman}},
  \bibinfo{journal}{Nature Mater} \textbf{\bibinfo{volume}{17}},
  \bibinfo{pages}{875} (\bibinfo{year}{2018}).

\bibitem[{\citenamefont{Beenakker and van
  Houten}(1991)}]{beenakker_josephson_1991}
\bibinfo{author}{\bibfnamefont{C.~W.~J.} \bibnamefont{Beenakker}}
  \bibnamefont{and} \bibinfo{author}{\bibfnamefont{H.}~\bibnamefont{van
  Houten}}, \bibinfo{journal}{Phys. Rev. Lett.} \textbf{\bibinfo{volume}{66}},
  \bibinfo{pages}{3056} (\bibinfo{year}{1991}), \bibinfo{note}{publisher:
  American Physical Society}.

\bibitem[{\citenamefont{Shabani et~al.}(2016)\citenamefont{Shabani, Kjaergaard,
  Suominen, Kim, Nichele, Pakrouski, Stankevic, Lutchyn, Krogstrup,
  Feidenhans'l et~al.}}]{shabani_two-dimensional_2016}
\bibinfo{author}{\bibfnamefont{J.}~\bibnamefont{Shabani}},
  \bibinfo{author}{\bibfnamefont{M.}~\bibnamefont{Kjaergaard}},
  \bibinfo{author}{\bibfnamefont{H.~J.} \bibnamefont{Suominen}},
  \bibinfo{author}{\bibfnamefont{Y.}~\bibnamefont{Kim}},
  \bibinfo{author}{\bibfnamefont{F.}~\bibnamefont{Nichele}},
  \bibinfo{author}{\bibfnamefont{K.}~\bibnamefont{Pakrouski}},
  \bibinfo{author}{\bibfnamefont{T.}~\bibnamefont{Stankevic}},
  \bibinfo{author}{\bibfnamefont{R.~M.} \bibnamefont{Lutchyn}},
  \bibinfo{author}{\bibfnamefont{P.}~\bibnamefont{Krogstrup}},
  \bibinfo{author}{\bibfnamefont{R.}~\bibnamefont{Feidenhans'l}},
  \bibnamefont{et~al.}, \bibinfo{journal}{Phys. Rev. B}
  \textbf{\bibinfo{volume}{93}}, \bibinfo{pages}{155402}
  (\bibinfo{year}{2016}).

\bibitem[{\citenamefont{Wickramasinghe
  et~al.}(2018)\citenamefont{Wickramasinghe, Mayer, Yuan, Nguyen, Jiao,
  Manucharyan, and Shabani}}]{wickramasinghe_transport_2018}
\bibinfo{author}{\bibfnamefont{K.~S.} \bibnamefont{Wickramasinghe}},
  \bibinfo{author}{\bibfnamefont{W.}~\bibnamefont{Mayer}},
  \bibinfo{author}{\bibfnamefont{J.}~\bibnamefont{Yuan}},
  \bibinfo{author}{\bibfnamefont{T.}~\bibnamefont{Nguyen}},
  \bibinfo{author}{\bibfnamefont{L.}~\bibnamefont{Jiao}},
  \bibinfo{author}{\bibfnamefont{V.}~\bibnamefont{Manucharyan}},
  \bibnamefont{and} \bibinfo{author}{\bibfnamefont{J.}~\bibnamefont{Shabani}},
  \bibinfo{journal}{Appl. Phys. Lett.} \textbf{\bibinfo{volume}{113}},
  \bibinfo{pages}{262104} (\bibinfo{year}{2018}).

\bibitem[{\citenamefont{Mayer et~al.}(2019)\citenamefont{Mayer, Yuan,
  Wickramasinghe, Nguyen, Dartiailh, and Shabani}}]{mayer_superconducting_2019}
\bibinfo{author}{\bibfnamefont{W.}~\bibnamefont{Mayer}},
  \bibinfo{author}{\bibfnamefont{J.}~\bibnamefont{Yuan}},
  \bibinfo{author}{\bibfnamefont{K.~S.} \bibnamefont{Wickramasinghe}},
  \bibinfo{author}{\bibfnamefont{T.}~\bibnamefont{Nguyen}},
  \bibinfo{author}{\bibfnamefont{M.~C.} \bibnamefont{Dartiailh}},
  \bibnamefont{and} \bibinfo{author}{\bibfnamefont{J.}~\bibnamefont{Shabani}},
  \bibinfo{journal}{Appl. Phys. Lett.} \textbf{\bibinfo{volume}{114}},
  \bibinfo{pages}{103104} (\bibinfo{year}{2019}).

\bibitem[{\citenamefont{Bøttcher et~al.}(2018)\citenamefont{Bøttcher,
  Nichele, Kjaergaard, Suominen, Shabani, Palmstrøm, and
  Marcus}}]{bottcher_superconducting_2018}
\bibinfo{author}{\bibfnamefont{C.~G.~L.} \bibnamefont{Bøttcher}},
  \bibinfo{author}{\bibfnamefont{F.}~\bibnamefont{Nichele}},
  \bibinfo{author}{\bibfnamefont{M.}~\bibnamefont{Kjaergaard}},
  \bibinfo{author}{\bibfnamefont{H.~J.} \bibnamefont{Suominen}},
  \bibinfo{author}{\bibfnamefont{J.}~\bibnamefont{Shabani}},
  \bibinfo{author}{\bibfnamefont{C.~J.} \bibnamefont{Palmstrøm}},
  \bibnamefont{and} \bibinfo{author}{\bibfnamefont{C.~M.}
  \bibnamefont{Marcus}}, \bibinfo{journal}{Nature Phys}
  \textbf{\bibinfo{volume}{14}}, \bibinfo{pages}{1138} (\bibinfo{year}{2018}),
  ISSN \bibinfo{issn}{1745-2481}, \bibinfo{note}{number: 11 Publisher: Nature
  Publishing Group}.

\bibitem[{\citenamefont{Mayer et~al.}(2020)\citenamefont{Mayer, Dartiailh,
  Yuan, Wickramasinghe, Rossi, and Shabani}}]{mayer_gate_2020}
\bibinfo{author}{\bibfnamefont{W.}~\bibnamefont{Mayer}},
  \bibinfo{author}{\bibfnamefont{M.~C.} \bibnamefont{Dartiailh}},
  \bibinfo{author}{\bibfnamefont{J.}~\bibnamefont{Yuan}},
  \bibinfo{author}{\bibfnamefont{K.~S.} \bibnamefont{Wickramasinghe}},
  \bibinfo{author}{\bibfnamefont{E.}~\bibnamefont{Rossi}}, \bibnamefont{and}
  \bibinfo{author}{\bibfnamefont{J.}~\bibnamefont{Shabani}},
  \bibinfo{journal}{Nat Commun} \textbf{\bibinfo{volume}{11}},
  \bibinfo{pages}{1} (\bibinfo{year}{2020}).

\bibitem[{\citenamefont{Fornieri et~al.}(2019)\citenamefont{Fornieri, Whiticar,
  Setiawan, Portolés, Drachmann, Keselman, Gronin, Thomas, Wang, Kallaher
  et~al.}}]{fornieri_evidence_2019}
\bibinfo{author}{\bibfnamefont{A.}~\bibnamefont{Fornieri}},
  \bibinfo{author}{\bibfnamefont{A.~M.} \bibnamefont{Whiticar}},
  \bibinfo{author}{\bibfnamefont{F.}~\bibnamefont{Setiawan}},
  \bibinfo{author}{\bibfnamefont{E.}~\bibnamefont{Portolés}},
  \bibinfo{author}{\bibfnamefont{A.~C.~C.} \bibnamefont{Drachmann}},
  \bibinfo{author}{\bibfnamefont{A.}~\bibnamefont{Keselman}},
  \bibinfo{author}{\bibfnamefont{S.}~\bibnamefont{Gronin}},
  \bibinfo{author}{\bibfnamefont{C.}~\bibnamefont{Thomas}},
  \bibinfo{author}{\bibfnamefont{T.}~\bibnamefont{Wang}},
  \bibinfo{author}{\bibfnamefont{R.}~\bibnamefont{Kallaher}},
  \bibnamefont{et~al.}, \bibinfo{journal}{Nature}
  \textbf{\bibinfo{volume}{569}}, \bibinfo{pages}{89} (\bibinfo{year}{2019}).

\bibitem[{\citenamefont{Casparis et~al.}(2019)\citenamefont{Casparis, Pearson,
  Kringhøj, Larsen, Kuemmeth, Nygård, Krogstrup, Petersson, and
  Marcus}}]{casparis_voltage-controlled_2019}
\bibinfo{author}{\bibfnamefont{L.}~\bibnamefont{Casparis}},
  \bibinfo{author}{\bibfnamefont{N.~J.} \bibnamefont{Pearson}},
  \bibinfo{author}{\bibfnamefont{A.}~\bibnamefont{Kringhøj}},
  \bibinfo{author}{\bibfnamefont{T.~W.} \bibnamefont{Larsen}},
  \bibinfo{author}{\bibfnamefont{F.}~\bibnamefont{Kuemmeth}},
  \bibinfo{author}{\bibfnamefont{J.}~\bibnamefont{Nygård}},
  \bibinfo{author}{\bibfnamefont{P.}~\bibnamefont{Krogstrup}},
  \bibinfo{author}{\bibfnamefont{K.~D.} \bibnamefont{Petersson}},
  \bibnamefont{and} \bibinfo{author}{\bibfnamefont{C.~M.}
  \bibnamefont{Marcus}}, \bibinfo{journal}{Phys. Rev. B}
  \textbf{\bibinfo{volume}{99}}, \bibinfo{pages}{085434}
  (\bibinfo{year}{2019}), \bibinfo{note}{publisher: American Physical Society}.

\bibitem[{\citenamefont{Courtois et~al.}(2008)\citenamefont{Courtois, Meschke,
  Peltonen, and Pekola}}]{courtois_origin_2008}
\bibinfo{author}{\bibfnamefont{H.}~\bibnamefont{Courtois}},
  \bibinfo{author}{\bibfnamefont{M.}~\bibnamefont{Meschke}},
  \bibinfo{author}{\bibfnamefont{J.~T.} \bibnamefont{Peltonen}},
  \bibnamefont{and} \bibinfo{author}{\bibfnamefont{J.~P.}
  \bibnamefont{Pekola}}, \bibinfo{journal}{Phys. Rev. Lett.}
  \textbf{\bibinfo{volume}{101}}, \bibinfo{pages}{067002}
  (\bibinfo{year}{2008}).

\bibitem[{\citenamefont{Gevorgian and Berg}(2001)}]{gevorgian_line_2001}
\bibinfo{author}{\bibfnamefont{S.}~\bibnamefont{Gevorgian}} \bibnamefont{and}
  \bibinfo{author}{\bibfnamefont{H.}~\bibnamefont{Berg}}, in
  \emph{\bibinfo{booktitle}{2001 31st {European} {Microwave} {Conference}}}
  (\bibinfo{year}{2001}), pp. \bibinfo{pages}{1--4}.

\bibitem[{\citenamefont{Niebler et~al.}(2009)\citenamefont{Niebler, Cuniberti,
  and Novotný}}]{niebler_analytical_2009}
\bibinfo{author}{\bibfnamefont{G.}~\bibnamefont{Niebler}},
  \bibinfo{author}{\bibfnamefont{G.}~\bibnamefont{Cuniberti}},
  \bibnamefont{and} \bibinfo{author}{\bibfnamefont{T.}~\bibnamefont{Novotný}},
  \bibinfo{journal}{Supercond. Sci. Technol.} \textbf{\bibinfo{volume}{22}},
  \bibinfo{pages}{085016} (\bibinfo{year}{2009}).

\bibitem[{\citenamefont{Yuan et~al.}(2019)\citenamefont{Yuan, Hatefipour,
  Magill, Mayer, Dartiailh, Sardashti, Wickramasinghe, Khodaparast, Matsuda,
  Kohama et~al.}}]{yuan_experimental_2019}
\bibinfo{author}{\bibfnamefont{J.}~\bibnamefont{Yuan}},
  \bibinfo{author}{\bibfnamefont{M.}~\bibnamefont{Hatefipour}},
  \bibinfo{author}{\bibfnamefont{B.~A.} \bibnamefont{Magill}},
  \bibinfo{author}{\bibfnamefont{W.}~\bibnamefont{Mayer}},
  \bibinfo{author}{\bibfnamefont{M.~C.} \bibnamefont{Dartiailh}},
  \bibinfo{author}{\bibfnamefont{K.}~\bibnamefont{Sardashti}},
  \bibinfo{author}{\bibfnamefont{K.~S.} \bibnamefont{Wickramasinghe}},
  \bibinfo{author}{\bibfnamefont{G.~A.} \bibnamefont{Khodaparast}},
  \bibinfo{author}{\bibfnamefont{Y.~H.} \bibnamefont{Matsuda}},
  \bibinfo{author}{\bibfnamefont{Y.}~\bibnamefont{Kohama}},
  \bibnamefont{et~al.}, \bibinfo{journal}{arXiv:1911.02738 [cond-mat]}
  (\bibinfo{year}{2019}), \bibinfo{note}{arXiv: 1911.02738}.

\bibitem[{\citenamefont{Suominen et~al.}(2017)\citenamefont{Suominen, Danon,
  Kjaergaard, Flensberg, Shabani, Palmstrøm, Nichele, and
  Marcus}}]{suominen_anomalous_2017}
\bibinfo{author}{\bibfnamefont{H.~J.} \bibnamefont{Suominen}},
  \bibinfo{author}{\bibfnamefont{J.}~\bibnamefont{Danon}},
  \bibinfo{author}{\bibfnamefont{M.}~\bibnamefont{Kjaergaard}},
  \bibinfo{author}{\bibfnamefont{K.}~\bibnamefont{Flensberg}},
  \bibinfo{author}{\bibfnamefont{J.}~\bibnamefont{Shabani}},
  \bibinfo{author}{\bibfnamefont{C.~J.} \bibnamefont{Palmstrøm}},
  \bibinfo{author}{\bibfnamefont{F.}~\bibnamefont{Nichele}}, \bibnamefont{and}
  \bibinfo{author}{\bibfnamefont{C.~M.} \bibnamefont{Marcus}},
  \bibinfo{journal}{Phys. Rev. B} \textbf{\bibinfo{volume}{95}},
  \bibinfo{pages}{035307} (\bibinfo{year}{2017}), \bibinfo{note}{publisher:
  American Physical Society}.

\bibitem[{\citenamefont{Rossignol et~al.}(2019)\citenamefont{Rossignol, Kloss,
  and Waintal}}]{Rossignol2019}
\bibinfo{author}{\bibfnamefont{B.}~\bibnamefont{Rossignol}},
  \bibinfo{author}{\bibfnamefont{T.}~\bibnamefont{Kloss}}, \bibnamefont{and}
  \bibinfo{author}{\bibfnamefont{X.}~\bibnamefont{Waintal}}, pp.
  \bibinfo{pages}{1--6} (\bibinfo{year}{2019}), \eprint{1901.05700}.

\bibitem[{\citenamefont{Averin and Bardas}(1995)}]{averin_bardas_1995}
\bibinfo{author}{\bibfnamefont{D.}~\bibnamefont{Averin}} \bibnamefont{and}
  \bibinfo{author}{\bibfnamefont{A.}~\bibnamefont{Bardas}},
  \bibinfo{journal}{Phys. Rev. Lett.} \textbf{\bibinfo{volume}{75}},
  \bibinfo{pages}{1831} (\bibinfo{year}{1995}).

\bibitem[{\citenamefont{English et~al.}(2016)\citenamefont{English, Hamilton,
  Chialvo, Moraru, Mason, and Van~Harlingen}}]{english_observation_2016}
\bibinfo{author}{\bibfnamefont{C.~D.} \bibnamefont{English}},
  \bibinfo{author}{\bibfnamefont{D.~R.} \bibnamefont{Hamilton}},
  \bibinfo{author}{\bibfnamefont{C.}~\bibnamefont{Chialvo}},
  \bibinfo{author}{\bibfnamefont{I.~C.} \bibnamefont{Moraru}},
  \bibinfo{author}{\bibfnamefont{N.}~\bibnamefont{Mason}}, \bibnamefont{and}
  \bibinfo{author}{\bibfnamefont{D.~J.} \bibnamefont{Van~Harlingen}},
  \bibinfo{journal}{Phys. Rev. B} \textbf{\bibinfo{volume}{94}},
  \bibinfo{pages}{115435} (\bibinfo{year}{2016}).

\bibitem[{\citenamefont{Sochnikov et~al.}(2015)\citenamefont{Sochnikov, Maier,
  Watson, Kirtley, Gould, Tkachov, Hankiewicz, Brüne, Buhmann, Molenkamp
  et~al.}}]{sochnikov_nonsinusoidal_2015}
\bibinfo{author}{\bibfnamefont{I.}~\bibnamefont{Sochnikov}},
  \bibinfo{author}{\bibfnamefont{L.}~\bibnamefont{Maier}},
  \bibinfo{author}{\bibfnamefont{C.~A.} \bibnamefont{Watson}},
  \bibinfo{author}{\bibfnamefont{J.~R.} \bibnamefont{Kirtley}},
  \bibinfo{author}{\bibfnamefont{C.}~\bibnamefont{Gould}},
  \bibinfo{author}{\bibfnamefont{G.}~\bibnamefont{Tkachov}},
  \bibinfo{author}{\bibfnamefont{E.~M.} \bibnamefont{Hankiewicz}},
  \bibinfo{author}{\bibfnamefont{C.}~\bibnamefont{Brüne}},
  \bibinfo{author}{\bibfnamefont{H.}~\bibnamefont{Buhmann}},
  \bibinfo{author}{\bibfnamefont{L.~W.} \bibnamefont{Molenkamp}},
  \bibnamefont{et~al.}, \bibinfo{journal}{Phys. Rev. Lett.}
  \textbf{\bibinfo{volume}{114}}, \bibinfo{pages}{066801}
  (\bibinfo{year}{2015}), \bibinfo{note}{publisher: American Physical Society}.

\bibitem[{\citenamefont{Kayyalha et~al.}(2020)\citenamefont{Kayyalha, Kazakov,
  Miotkowski, Khlebnikov, Rokhinson, and Chen}}]{kayyalha_highly_2020}
\bibinfo{author}{\bibfnamefont{M.}~\bibnamefont{Kayyalha}},
  \bibinfo{author}{\bibfnamefont{A.}~\bibnamefont{Kazakov}},
  \bibinfo{author}{\bibfnamefont{I.}~\bibnamefont{Miotkowski}},
  \bibinfo{author}{\bibfnamefont{S.}~\bibnamefont{Khlebnikov}},
  \bibinfo{author}{\bibfnamefont{L.~P.} \bibnamefont{Rokhinson}},
  \bibnamefont{and} \bibinfo{author}{\bibfnamefont{Y.~P.} \bibnamefont{Chen}},
  \bibinfo{journal}{npj Quantum Mater.} \textbf{\bibinfo{volume}{5}},
  \bibinfo{pages}{1} (\bibinfo{year}{2020}), ISSN \bibinfo{issn}{2397-4648},
  \bibinfo{note}{number: 1 Publisher: Nature Publishing Group}.

\bibitem[{\citenamefont{Spanton et~al.}(2017)\citenamefont{Spanton, Deng,
  Vaitiekėnas, Krogstrup, Nygård, Marcus, and
  Moler}}]{spanton_currentphase_2017}
\bibinfo{author}{\bibfnamefont{E.~M.} \bibnamefont{Spanton}},
  \bibinfo{author}{\bibfnamefont{M.}~\bibnamefont{Deng}},
  \bibinfo{author}{\bibfnamefont{S.}~\bibnamefont{Vaitiekėnas}},
  \bibinfo{author}{\bibfnamefont{P.}~\bibnamefont{Krogstrup}},
  \bibinfo{author}{\bibfnamefont{J.}~\bibnamefont{Nygård}},
  \bibinfo{author}{\bibfnamefont{C.~M.} \bibnamefont{Marcus}},
  \bibnamefont{and} \bibinfo{author}{\bibfnamefont{K.~A.} \bibnamefont{Moler}},
  \bibinfo{journal}{Nature Phys} \textbf{\bibinfo{volume}{13}},
  \bibinfo{pages}{1177} (\bibinfo{year}{2017}), ISSN \bibinfo{issn}{1745-2481},
  \bibinfo{note}{number: 12 Publisher: Nature Publishing Group}.

\bibitem[{\citenamefont{Askerzade}(2015)}]{askerzade_effects_2015}
\bibinfo{author}{\bibfnamefont{I.~N.} \bibnamefont{Askerzade}},
  \bibinfo{journal}{Low Temperature Physics} \textbf{\bibinfo{volume}{41}},
  \bibinfo{pages}{241} (\bibinfo{year}{2015}), ISSN \bibinfo{issn}{1063-777X},
  \bibinfo{note}{publisher: American Institute of Physics}.

\bibitem[{\citenamefont{Snyder et~al.}(2018)\citenamefont{Snyder, Trimble,
  Rong, Folkes, Taylor, and Williams}}]{snyder_weak-link_2018}
\bibinfo{author}{\bibfnamefont{R.}~\bibnamefont{Snyder}},
  \bibinfo{author}{\bibfnamefont{C.}~\bibnamefont{Trimble}},
  \bibinfo{author}{\bibfnamefont{C.}~\bibnamefont{Rong}},
  \bibinfo{author}{\bibfnamefont{P.}~\bibnamefont{Folkes}},
  \bibinfo{author}{\bibfnamefont{P.}~\bibnamefont{Taylor}}, \bibnamefont{and}
  \bibinfo{author}{\bibfnamefont{J.}~\bibnamefont{Williams}},
  \bibinfo{journal}{Phys. Rev. Lett.} \textbf{\bibinfo{volume}{121}},
  \bibinfo{pages}{097701} (\bibinfo{year}{2018}).

\bibitem[{\citenamefont{Lee et~al.}(2015)\citenamefont{Lee, Kim, Jhi, and
  Lee}}]{lee_ultimately_2015}
\bibinfo{author}{\bibfnamefont{G.-H.} \bibnamefont{Lee}},
  \bibinfo{author}{\bibfnamefont{S.}~\bibnamefont{Kim}},
  \bibinfo{author}{\bibfnamefont{S.-H.} \bibnamefont{Jhi}}, \bibnamefont{and}
  \bibinfo{author}{\bibfnamefont{H.-J.} \bibnamefont{Lee}},
  \bibinfo{journal}{Nat Commun} \textbf{\bibinfo{volume}{6}},
  \bibinfo{pages}{1} (\bibinfo{year}{2015}), ISSN \bibinfo{issn}{2041-1723},
  \bibinfo{note}{number: 1 Publisher: Nature Publishing Group}.

\bibitem[{\citenamefont{Panghotra et~al.}(2020)\citenamefont{Panghotra, Raes,
  Silva, Cools, Keijers, Scheerder, Moshchalkov, and
  Vondel}}]{panghotra_giant_2020}
\bibinfo{author}{\bibfnamefont{R.}~\bibnamefont{Panghotra}},
  \bibinfo{author}{\bibfnamefont{B.}~\bibnamefont{Raes}},
  \bibinfo{author}{\bibfnamefont{C.~C. d.~S.} \bibnamefont{Silva}},
  \bibinfo{author}{\bibfnamefont{I.}~\bibnamefont{Cools}},
  \bibinfo{author}{\bibfnamefont{W.}~\bibnamefont{Keijers}},
  \bibinfo{author}{\bibfnamefont{J.~E.} \bibnamefont{Scheerder}},
  \bibinfo{author}{\bibfnamefont{V.~V.} \bibnamefont{Moshchalkov}},
  \bibnamefont{and} \bibinfo{author}{\bibfnamefont{J.~V.~d.}
  \bibnamefont{Vondel}}, \bibinfo{journal}{Commun Phys}
  \textbf{\bibinfo{volume}{3}}, \bibinfo{pages}{1} (\bibinfo{year}{2020}).

\end{thebibliography}

\begin{thebibliography}{5}
\makeatletter
\addtocounter{NAT@ctr}{50}
\makeatother
\expandafter\ifx\csname natexlab\endcsname\relax\def\natexlab#1{#1}\fi
\expandafter\ifx\csname bibnamefont\endcsname\relax
  \def\bibnamefont#1{#1}\fi
\expandafter\ifx\csname bibfnamefont\endcsname\relax
  \def\bibfnamefont#1{#1}\fi
\expandafter\ifx\csname citenamefont\endcsname\relax
  \def\citenamefont#1{#1}\fi
\expandafter\ifx\csname url\endcsname\relax
  \def\url#1{\texttt{#1}}\fi
\expandafter\ifx\csname urlprefix\endcsname\relax\def\urlprefix{URL }\fi
\providecommand{\bibinfo}[2]{#2}
\providecommand{\eprint}[2][]{\url{#2}}

\bibitem[{\citenamefont{Russer}(1972)}]{Russer_1972}
\bibinfo{author}{\bibfnamefont{P.}~\bibnamefont{Russer}},
  \bibinfo{journal}{Journal of Applied Physics} \textbf{\bibinfo{volume}{43}},
  \bibinfo{pages}{2008} (\bibinfo{year}{1972}).

\bibitem[{\citenamefont{Dorokhov}(1984)}]{Dorokhov_1984}
\bibinfo{author}{\bibfnamefont{O.}~\bibnamefont{Dorokhov}},
  \bibinfo{journal}{Solid State Communications} \textbf{\bibinfo{volume}{51}},
  \bibinfo{pages}{381 } (\bibinfo{year}{1984}), ISSN \bibinfo{issn}{0038-1098}.

\bibitem[{\citenamefont{van Rossum and
  Nieuwenhuizen}(1999)}]{Rossum_1999_review}
\bibinfo{author}{\bibfnamefont{M.~C.~W.} \bibnamefont{van Rossum}}
  \bibnamefont{and} \bibinfo{author}{\bibfnamefont{T.~M.}
  \bibnamefont{Nieuwenhuizen}}, \bibinfo{journal}{Rev. Mod. Phys.}
  \textbf{\bibinfo{volume}{71}}, \bibinfo{pages}{313} (\bibinfo{year}{1999}).

\bibitem[{\citenamefont{Domínguez et~al.}(2012)\citenamefont{Domínguez,
  Hassler, and Platero}}]{dominguez_dynamical_2012}
\bibinfo{author}{\bibfnamefont{F.}~\bibnamefont{Domínguez}},
  \bibinfo{author}{\bibfnamefont{F.}~\bibnamefont{Hassler}}, \bibnamefont{and}
  \bibinfo{author}{\bibfnamefont{G.}~\bibnamefont{Platero}},
  \bibinfo{journal}{Phys. Rev. B} \textbf{\bibinfo{volume}{86}},
  \bibinfo{pages}{140503} (\bibinfo{year}{2012}).

\end{thebibliography}

\renewcommand{\thefigure}{\textbf{S\arabic{figure}}}
\setcounter{figure}{0}
\setcounter{equation}{0} 

\clearpage

\section*{Supplementary information}

\section{Fabrication and measuremnts methods}

For both samples, the mesa is 4$\mu$m wide. The fabrication process is performed by electron beam lithography using PMMA resist. Transene type D is used for wet etching the Al and a III-V wet etch ($H_2O$ : $C_6H_8O_7$ : $H_3PO_4$ : $H_2O_2$) to define deep semiconductor mesas. 

Devices are measured in a 4-point geometry using a current bias configuration. Differential resistance is measured using a lock-in amplifier SRS860. Microwave excitation is provided through a nearby antenna. All measurements are performed in a dilution fridge with mixing chamber temperature of 30 mK.

\section{Shapiro steps at lower frequency}

\begin{figure}[ht!]
\centering
\includegraphics[width=0.48\textwidth]{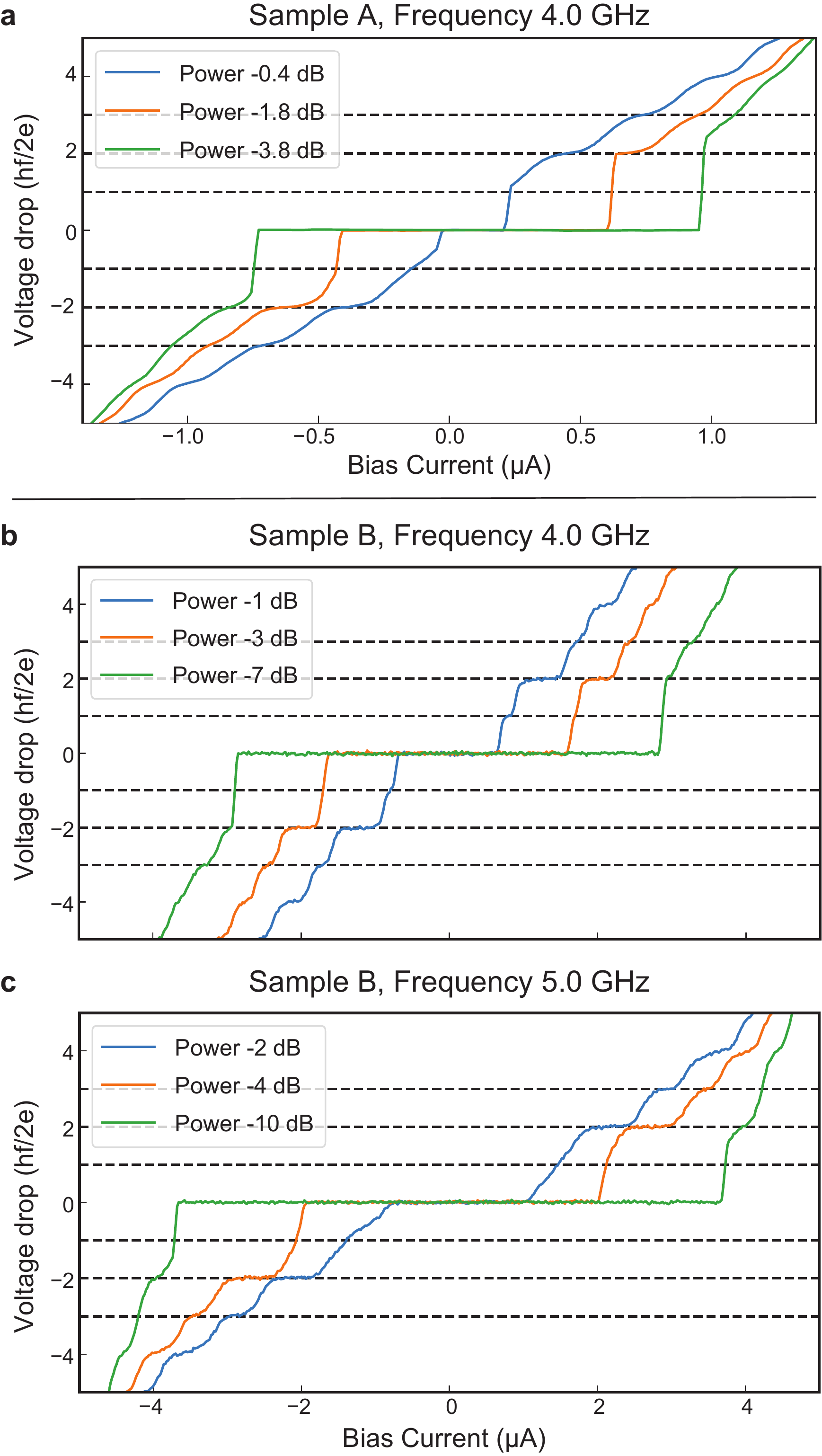}
\caption{\label{fig-s1}\textbf{VI characteristic at lower frequency}. \textbf{a}, Sample A at 4 GHz, \textbf{b, c}, Sample B at 4 and 5 GHz.}
\end{figure}

Figure \ref{fig-s1} presents data on both sample A and B at the lowest frequency achievable in our setup. This limit is mostly due to the weak coupling of the microwave antenna to the sample at low frequency.

On both samples, one can observe the same missing first step described in the main text. In addition, on sample B the third step appears partially suppressed which is consistent with data presented at 6 GHz in the main text.

\section{Semiclassical Treatment of a Current-driven Josephson Junction}

    The dynamics of a current-driven JJ can be treated within a semiclassical description where the current is carried in three parallel channels: a resistive channel describing dissipative current, a capacitive channel describing charge accumulation on the superconducting leads across the weak link, and a supercurrent channel describing current mediated by Cooper pairs. Our JJ has a small geometric capacitance ($C \sim 1$ fF) corresponding to the overdamped regime; thus, we neglect the capacitive channel. Driving our JJ with an ac current, we can use Kirchoff's junction law to write
    \begin{equation}
        I_{drive}(t) = I_R(\dot{\phi}) + I_s(\phi)
    \end{equation}
    where $\phi$ is the phase difference between the two superconducting electrodes, $I_{drive}(t) = I_{dc} + I_{ac} \sin (2\pi f_{ac}t)$ is the external driving current, $I_{R}(\dot{\phi}) = \frac{\hbar}{2 e R}\dot{\phi}$ is the current in the resistive channel, and $I_s(\phi)$ is the supercurrent contribution. Then the master equation governing our JJ dynamics is
    \begin{equation}
        \frac{\hbar}{2 e I_c R} \dot{\phi} = \hat{I}_{dc} + \hat{I}_{ac} \sin(2 \pi f_{ac} t) - \hat{I}_s(\phi),
        \label{EqMotion-sup}
    \end{equation}
    where the hats denote currents normalized by the critical supercurrent $I_c$. Numerical results for an ideal current bias have been discussed before for a topologically-trivial JJ \cite{Russer_1972}. 

\subsection{Bimodal Distribution Model}

    In JJ's, the supercurrent is mediated by Andreev bound states (ABS) with energy given by 
    \begin{equation}
        E_{ABS} = \pm \Delta \sqrt{1 - \tau \sin^2(\phi/2)}
        \label{ABSenergy}
    \end{equation}
    where $\Delta$ is the SC gap and $\tau$ is the transparency of the ABS. The spectrum according to Eq. \ref{ABSenergy} exhibits a gap $2\Delta \sqrt{1-\tau}$ at $\phi = \pi$, which is an avoided crossing due to the broken translational symmetry at the interface between the weak link and the superconducting electrodes. The supercurrent carried by a single ABS at zero temperature is given by a skewed sinusoidal CPR,
    \begin{equation}
        I_{ABS} = \frac{e\Delta}{2 \hbar} \frac{\tau \sin(\phi)}{\sqrt{1 - \tau \sin^2(\phi/2)}}.
    \end{equation}

    In what follows, we will assume the subgap modes in the JJ (which dominate the contribution to the supercurrent) have a distribution of transparencies according to a bimodal distribution. Bimodal distributions have been discussed before in the context of diffusive junctions \cite{Dorokhov_1984, Rossum_1999_review}. We consider two \textit{effective} modes: a low-transparency mode with a sinusoidal CPR and a high-transparency mode with a skewed CPR determined by an \textit{effective} transparency $\tau$. Then we write our supercurrent as
    \begin{equation}
        I_s = I_c \left( \frac{n}{\alpha_0} \sin(\phi) + \frac{1-n}{\alpha_{\tau}} \frac{\sin(\phi)}{\sqrt{1 - \tau\sin^2(\phi/2)}} \right),
    \end{equation}
    where $n$ is the fraction of critical current contributed by the effective low-transparency mode and $I_c$ is the critical current. We have included normalizations $\alpha_0$ and $\alpha_{\tau}$ that are determined by,
    \begin{equation}
        \alpha_0 = \frac{1}{\sin(\tilde{\phi}_{max})}, \quad \alpha_{\tau} = \frac{\sqrt{1 - \tau \sin^2(\tilde{\phi}_{max})}}{\sin(\tilde{\phi}_{max})}
    \end{equation}
    where $\tilde{\phi}_{max}$ is such that 
    \begin{gather*}
        max\left(  n \sin(\phi) + (1-n) \frac{\sin(\phi)}{\sqrt{1 - \tau\sin^2(\phi/2)}}  \right)\\
        = n \sin(\tilde{\phi}_{max}) + (1-n) \frac{\sin(\tilde{\phi}_{max})}{\sqrt{1 - \tau\sin^2(\tilde{\phi}_{max}/2)}} 
    \end{gather*}

\subsection{Landau-Zener Processes}

    In a two-level quantum system, Landau-Zener processes describe diabatic energy level transitions. Generally, the Landau-Zener transition (LZT) probability will depend on the difference in energy between the two states and the rate at which the dynamical variable changes: a small energy gap and rapid evolution of the dynamical variable are favorable conditions for an LZT to occur. We can treat the ground state and excited state of a single ABS with transparency $\tau$ as a two-level quantum system and solve for the LZT probability at the avoided crossing \cite{averin_bardas_1995}:
    \begin{equation}
         P_{LZT}(t) = \exp\left( - \pi\,\frac{\Delta (1-\tau)}{e\,\vert V(t)\vert} \right).
    \end{equation}
    Here we neglect interference effects due to phase fluctuations and coherence between LZT's. 

    A successful LZT will change the sign of the supercurrent contribution due to the ABS mode undergoing the transition. For high transparency modes, LZT probability can be significant because of the small gap at the avoided crossing. We model the collective behavior of the high-transparency modes by considering a single \textit{effective} LZT in our calculations occurring at avoided crossings. Thus, we take our supercurrent to be given by,
    \begin{equation}
        I_s = I_c \left( \frac{n}{\alpha_0} \sin(\phi) + s\frac{1-n}{\alpha_{\tau}} \frac{\sin(\phi)}{\sqrt{1 - \tau\sin^2(\phi/2)}} \right),
        \label{eq:Is-sup}
    \end{equation}
    where $s=\pm$ controls the sign flip due to an LZT. Following Ref. \cite{dominguez_dynamical_2012}, we solve Eq. \ref{EqMotion-sup} dynamically to account for LZT's at the avoided crossing of the effective high-transparency mode.

\begin{figure}[ht!]
        \centering
        \includegraphics[width=0.48\textwidth]{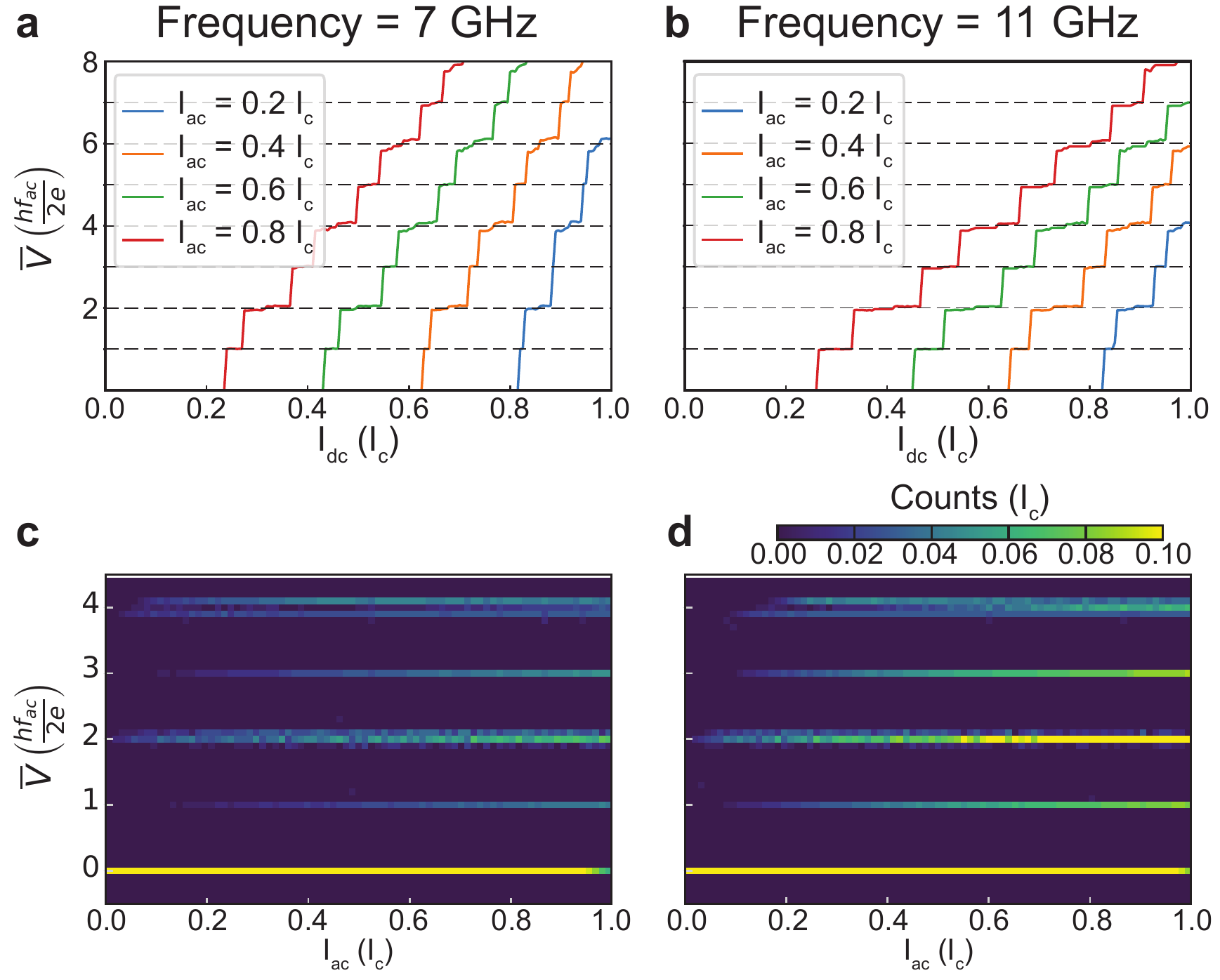}
        \caption{\label{fig:sampA-VI} (a,b) VI characteristics for various ac current biases with sample A parameters. (c,d) Histogram of Josephson junction voltage as a function of power at a fixed ac driving frequency.}
    \end{figure}

\begin{figure}[ht!]
        \centering
        \includegraphics[width=0.48\textwidth]{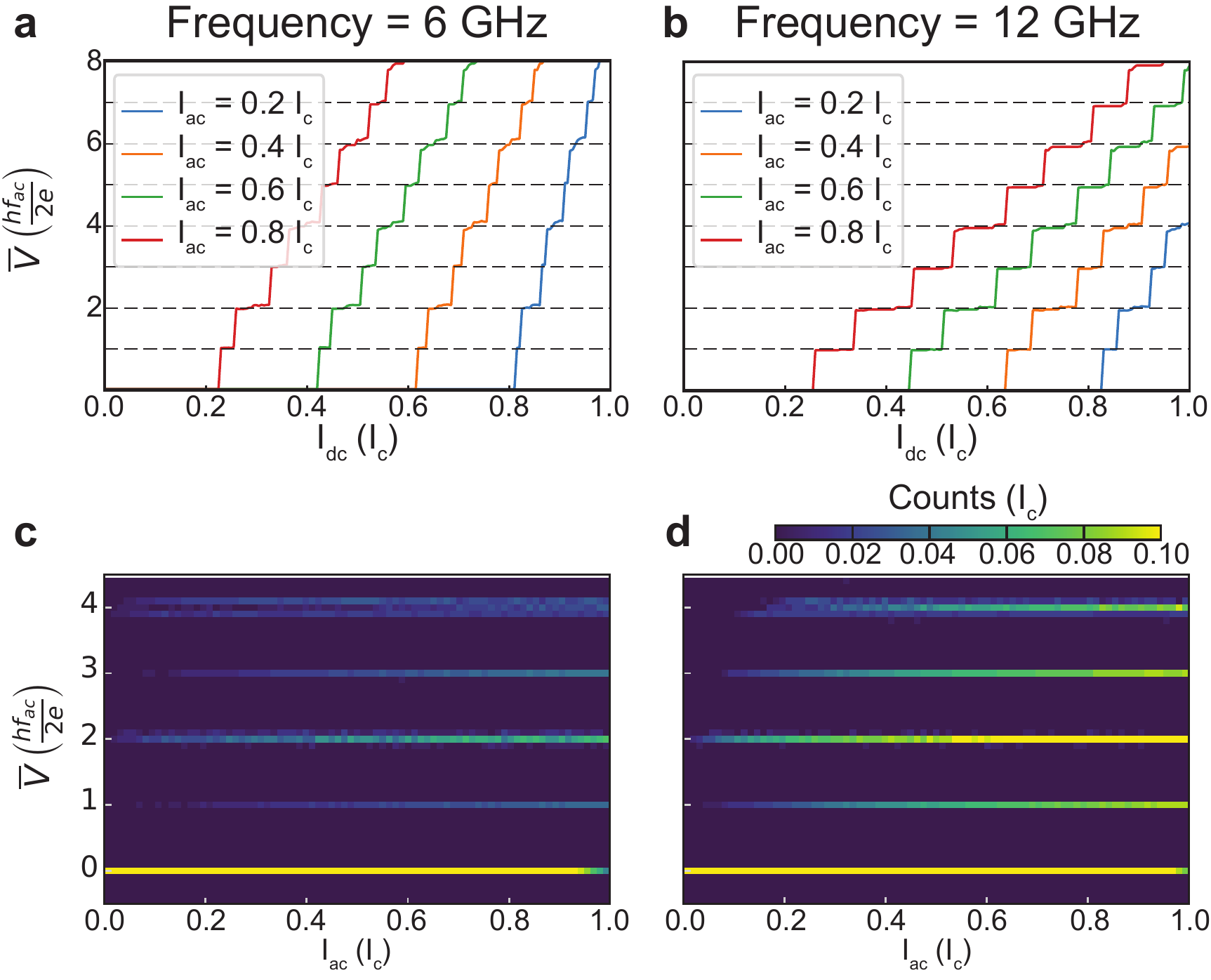}
        \caption{\label{fig:sampB-VI} (a,b) VI characteristics for various ac current biases with sample B parameters. (c,d) Histogram of Josephson junction voltage as a function of power at a fixed ac driving frequency.}
    \end{figure}

    Fig. \ref{fig:sampA-VI} and \ref{fig:sampB-VI} shows the main results from the main text along with histograms of Josephson junction voltage as function of ac power. We observe odd steps gradually suppressed at low driving frequencies and low power which qualitatively agrees with experimental results. 

\subsection{V(t) in BMD Model}

    \begin{figure}[ht!]
        \centering
        \includegraphics[width=0.48\textwidth]{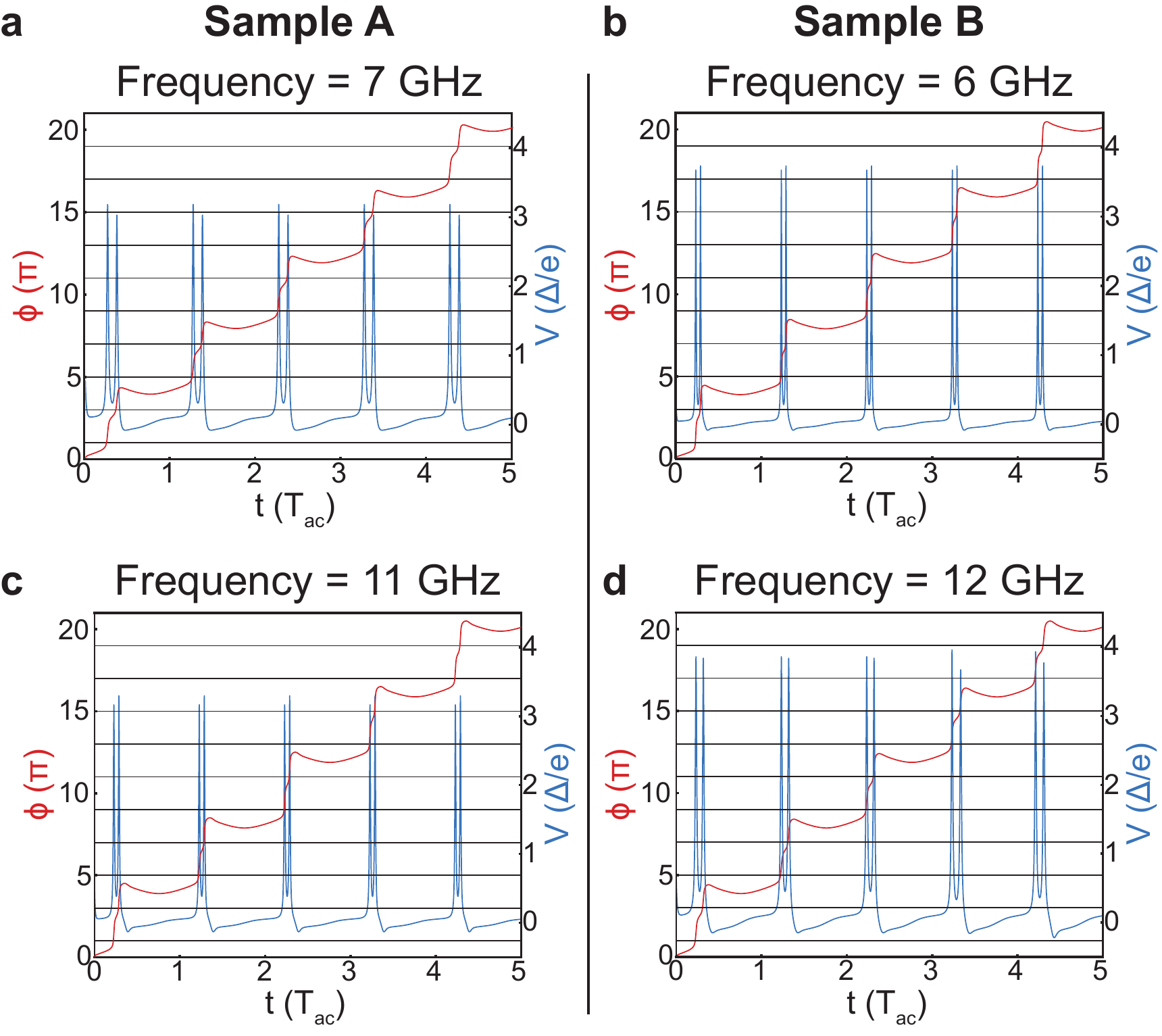}
        \caption{\label{fig:instV} Phase and instantaneous voltage across the JJ as function of time. Simulations for sample A using $\tau=0.98$ and $n=0.95$ at power $I_1 = 0.4I_c$ for (a) frequency $7$ GHz and $I_0 = 0.7I_c$, and (c) frequency $11$ GHz and $I_0 = 0.8I_c$. Simulations for sample B using $\tau=0.98$ and $n=0.97$ at power $I_1 = 0.4I_c$ for (a) frequency $6$ GHz and $I_0 = 0.7I_c$, and (c) frequency $11$ GHz and $I_0 = 0.8I_c$.}
    \end{figure}

    The time evolution of $\phi$ and induced voltage (proportional to $\dot{\phi}$) across the junction for $I_0\approx 0.7I_c$ and $I_1 = 0.4I_c$ are shown in Fig.~\ref{fig:instV}. Notice that the induced voltage has a negligible dependence on ac driving frequency during the cross-over from all steps visible to suppressed odd steps. This seems to be a robust feature for both parameters sets at powers and dc current biases below $I_c$. 
    
\begin{figure}[ht!]
        \centering
        \includegraphics[width=0.48\textwidth]{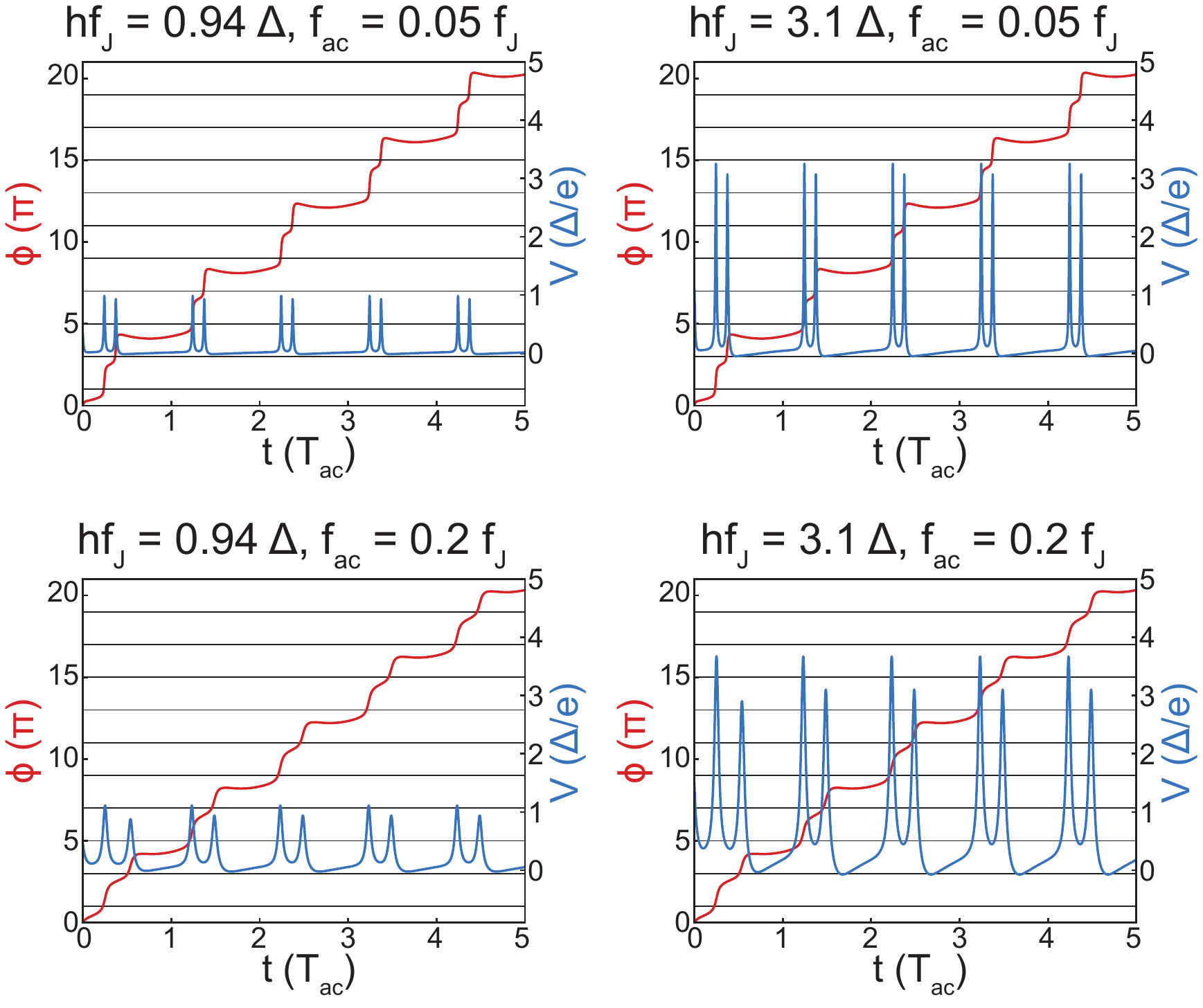}
        \caption{\label{fig:pureSin} Phase and instantaneous voltage across the JJ as function of time. We only consider the sinusoidal part of Eq. \ref{eq:Is-sup} ($n=1$) for various Josephson frequencies and driving frequencies. }
    \end{figure}

    To illustrate this, we can start with a simplified picture where we only consider a purely sinusoidal CPR in the absence of LZ transitions (i.e. $n=1$). Phase and instantaneous voltage across the JJ for $f_J = 3.1\Delta$ (corresponding to $I_cR_n$ for sample A) and $f_J = 0.94\Delta$ for driving frequencies $f_ac = 0.2f_J$ and $0.05f_J$ are shown in Fig. \ref{fig:pureSin}. We notice that the peaks of the resonances in $V$ correspond closely to $hf_J/e$ and are very weakly dependent on driving frequency. Numerically, we observe that these dependencies survive when we include a skewed CPR and LZ transitions.

\subsection{Landau-Zener Transitions in BMD Model}

    \begin{figure}[ht!]
        \centering
        \includegraphics[width=0.48\textwidth]{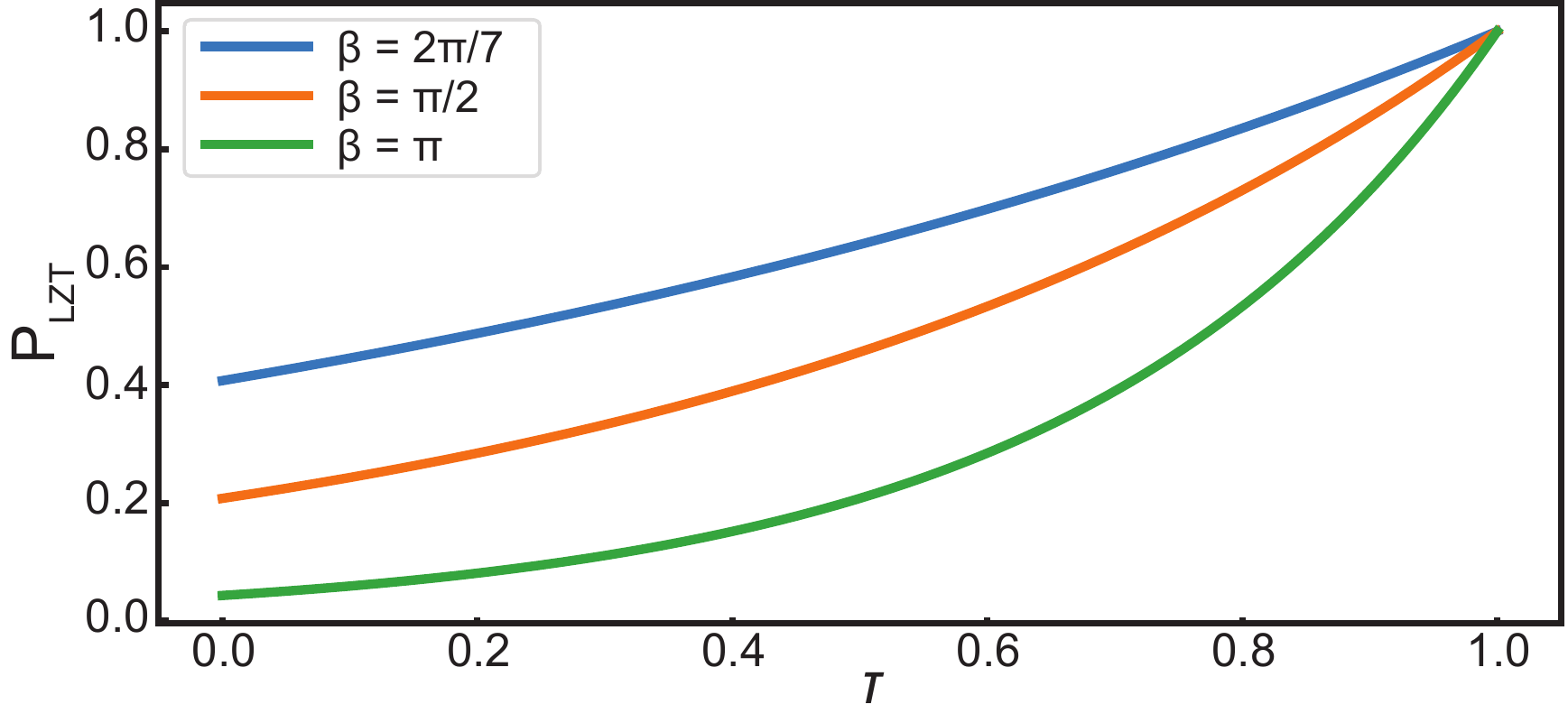}
        \caption{\label{fig:PrLZT} Probability of an LZ transition as a function of transparency $\tau$ for fixed $\beta = \frac{\pi \Delta}{e\vert V\vert}$. The value of $\beta$ was chosen based on typical values for $\vert V\vert$ at $\phi=\pi$ where a transition is allowed in the simulations.}
    \end{figure}
    
    Ignoring the fluctuations in the value of $\vert \dot{\phi} \vert$ as $\phi$ sweeps through odd integer multiples of $\pi$, we can consider
    \begin{equation}
        P_{LZT} = \exp \left( -\beta (1-\tau) \right)
    \end{equation}
    where $\frac{2\pi}{7} < \beta < \pi$ is a constant. We have chosen this range of $\beta$ based on possible values of $V$ giving rise to an LZ transition in our simulations. Results for various values of $\beta$ are shown in Fig.~\ref{fig:PrLZT}.
    
    \begin{figure}[ht!]
        \centering
        \includegraphics[width=0.48\textwidth]{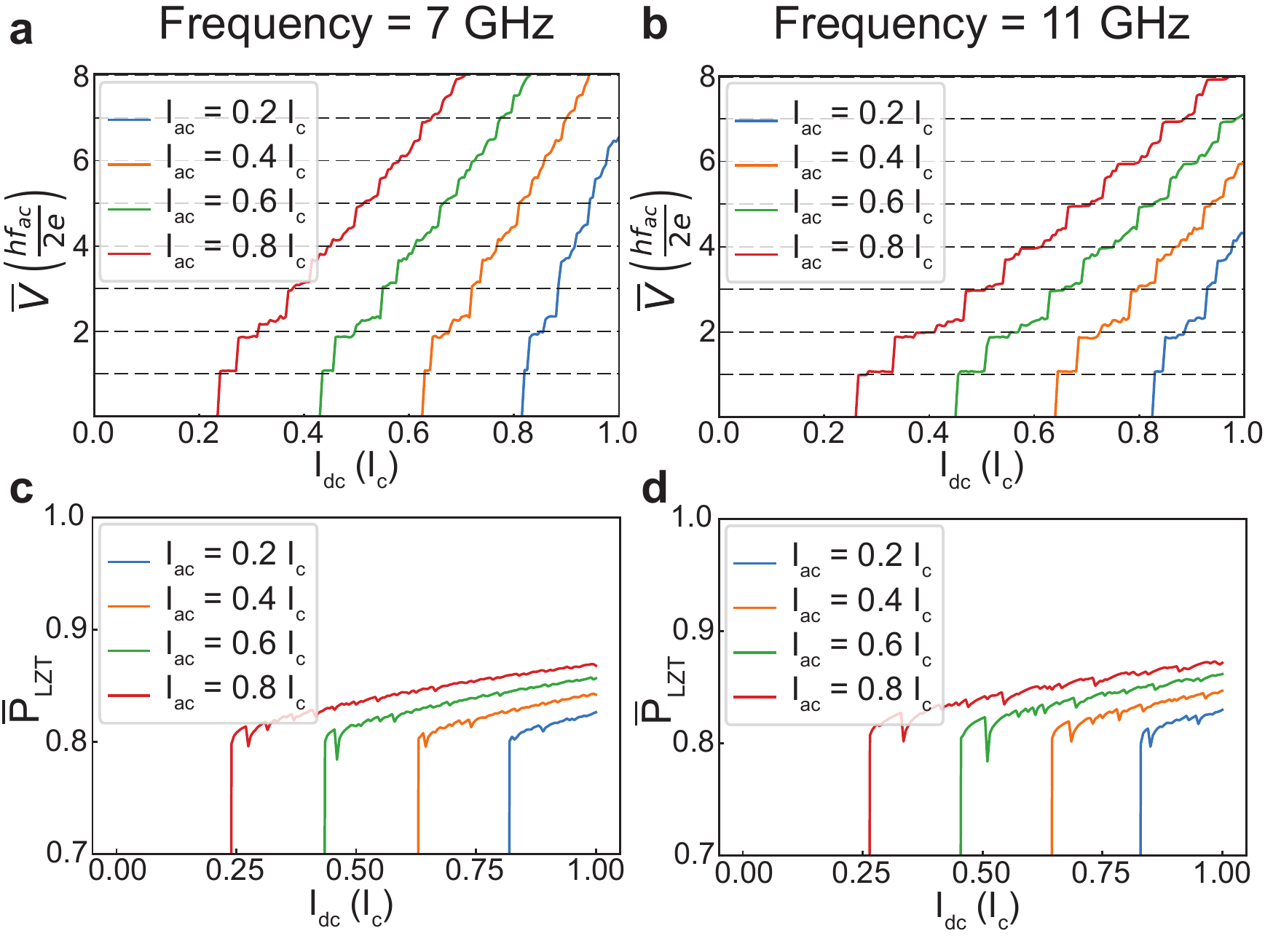}
        \caption{\label{fig:moderateT} (a,b) VI characteristics for sample A using a lower transparency $\tau=0.9$ and $n=0.95$. Simulations at lower transparency show deteriorated quantized steps due to a reduced probability of LZ transitions. (c,d) Time-averaged probability of LZ transitions as the phase passes an avoided crossing. Clearly, the probability of an LZ transition must be very near unity to observe sharp quantized steps.}
    \end{figure}

    For transparency $<0.98$, the LZT probability takes on values $<0.97$ leading to departures from quantized values of induced voltage--integer multiples of $\frac{h f_{ac}}{2e}$ \cite{dominguez_dynamical_2012}. Fig. \ref{fig:moderateT} shows VI curves for $\tau=0.9$ and $n=0.95$ using sample A values for $I_c$ and $R_n$. Results are similar for sample B. Clearly Shapiro steps are not well defined, and even-integer steps generally show small steps about the quantized values $\frac{h f_{ac}}{2e}$. The experimental samples have very robust, quantized steps implying only high transparency modes effectively participate in Landau-Zener processes.

\end{document}